\documentclass[
reprint,
superscriptaddress,
amsmath,
amssymb,
aps,
longbibliography,
pra,
showpacs,
floatfix
]{revtex4-1}
\usepackage{graphicx}
\usepackage{xcolor} 
\usepackage{soul,color}
\usepackage{helvet}
\usepackage{array}
\usepackage{dcolumn}

\newcolumntype{K}[1]{>{\centering\arraybackslash}p{#1}}

\usepackage{multirow}
\usepackage{braket}
\usepackage{comment}
\usepackage{float}

\usepackage[colorlinks=true,linkcolor=blue,citecolor=blue,urlcolor=blue]{hyperref}

\begin{document}

\title{Electro-phononic and magneto-phononic frequency conversion}

\author{Daniel\ A.\ Bustamante Lopez}
\email{dabl@bu.edu}
\affiliation{Department of Physics, Boston University, Boston, Massachusetts 02215, USA}
\author{Wanzheng\ Hu}
\affiliation{Department of Physics, Boston University, Boston, Massachusetts 02215, USA}
\affiliation{Division of Materials Science and Engineering, Boston University, Boston, Massachusetts 02215, USA}
\affiliation{Photonics Center, Boston University, Boston, Massachusetts 02215, USA}
\author{Dominik~M.\ Juraschek}
\email{djuraschek@tauex.tau.ac.il}
\affiliation{School of Physics and Astronomy, Tel Aviv University, Tel Aviv 6997801, Israel}

\date{\today}



\begin{abstract}
Nonlinear frequency conversion by optical rectification, as well as difference- and sum-frequency generation are fundamental processes for producing electromagnetic radiation at different frequencies. Here, we demonstrate that coherently excited infrared-active phonons can be used as transducers for generating nonlinear electric polarizations and magnetizations via phonon-phonon and phonon-magnon interactions, in a way similar to nonlinear optical frequency conversion. We derive analytical solutions for the time-dependent polarizations and magnetizations for the second-order response to the electric field component of an ultrashort laser pulse. These allow us to define second-order nonlinear electric and magneto-electric susceptibilities that capture the rectification, as well as the impulsive and sum-frequency excitation of coherent phonons and magnons. Our theoretical framework naturally incorporates existing mechanisms and further leads to the prediction of a hybrid magneto-opto-phononic inverse Faraday effect involving photon-phonon-magnon scattering. Our work demonstrates nonlinear phononics as a pathway to controlling the electric polarization and magnetization in solids.
\end{abstract}

\maketitle


\section{Introduction}

Ultrafast pump-probe spectroscopy has enabled the study of collective excitations in solids on fundamental timescales, such as lattice, spin, and charge dynamics \cite{Basov2017,Disa2021}. Nonlinear optical frequency-conversion processes, such as optical rectification, difference-frequency generation, and sum-frequency generation hereby play a crucial role, generating light across various spectral ranges, and serving as a probe for the collective excitations and electronic phases of materials. For example, optical rectification and difference-frequency generation are some of the primary methods to generate terahertz radiation today \cite{Dhillon2017}. Further, time-resolved second-harmonic generation, a special case of sum-frequency generation, has been used as a sensitive probe for detecting magnetic and ferroelectric properties and dynamics in recent years \cite{Mankowsky_2:2017,Li2019,Nova2019,Tzschaschel2019,Henstridge2022,Toyoda2021,BustamanteLopez2023}. 

\begin{figure*}[t]
\centering
\includegraphics[width=0.8\linewidth]{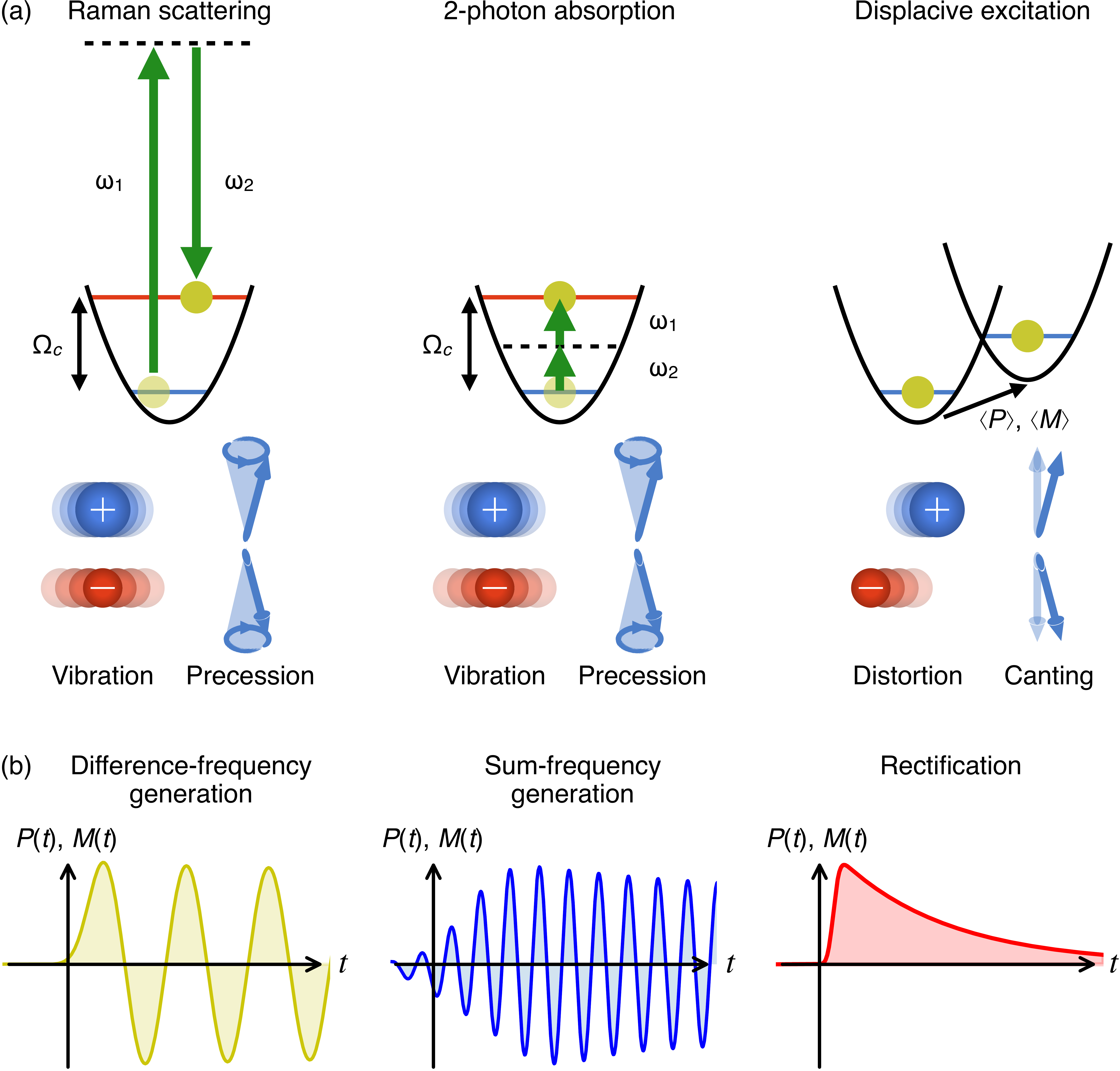}
\caption{
\textbf{Overview of electro-phononic and magneto-phononic mechanisms.} (a) Three mechanisms for the excitation of phonons or magnons with an eigenfrequency $\Omega_c$, driven by the electric field component of an ultrashort laser pulse with frequency components $\omega_1$ and $\omega_2$. Left panel, Raman scattering impulsively generates coherent phonons or magnons. Middle panel, sum-frequency excitation non-impulsively generates coherent phonons or magnons in a two-particle absorption process. If $\omega_1=\omega_2$, sum-frequency generation becomes second-harmonic generation. Both cases result in vibrational motions of the atoms or spin precessions, respectively. Right panel, displacive excitation quasistatically displaces the collective mode's coordinate, which induces a structural distortion or spin canting. (b) Schematic time evolution of the amplitudes of vibration or precession of the phonon or magnon modes in the above three cases. A Raman-active phonon in a noncentrosymmetric material induces an electric polarization, $P(t)$, whereas a magnon in an antiferromagnetic material induces a magnetization, $M(t)$.
}
\label{fig:mechanisms}
\end{figure*}

At the same time, these processes can be used to coherently excite collective modes, in particular lattice vibrations (phonons) and spin waves (magnons), as depicted in Fig.~\ref{fig:mechanisms}a. In impulsive stimulated Raman scattering (ISRS), the difference frequency of two photons excites a Raman-active phonon or magnon \cite{desilvestri:1985,Kalashnikova2007}, whereas in two-photon absorption, the sum frequency of two photons does so \cite{maehrlein:2017,Juraschek2018}. Both mechanisms induce coherent vibrational motions or spin precessions (Fig.~\ref{fig:mechanisms}a). In addition, laser pulses with photon energies above the band gap of the material create electronic excitations that change the interatomic potential energy landscape. This mechanism induces unidirectional displacements of the collective mode coordinate that lead to a structural distortion or spin canting (Fig.~\ref{fig:mechanisms}a), also known as displacive excitation of coherent phonons or magnons (DECP/DECM) \cite{zeiger:1992,Kalashnikova2008,Giorgianni2022}. The time evolution of the amplitudes of the atomic vibrations or spin precessions is schematically shown for each case in Fig.~\ref{fig:mechanisms}b.

In recent years, an increasing amount of Raman scattering \cite{forst:2011,Forst2013,mankowsky:2014,subedi:2014,subedi:2015,Mankowsky:2015,fechner:2016,Mankowsky:2017,juraschek:2017,Gu2017,Gu2018,Khalsa2018,Kaaret2022,Henstridge2022,Khalsa2023,Caruso2023,Neugebauer2021,nova:2017,Afanasiev2021} and two-photon absorption processes \cite{Juraschek2018,Knighton2019,Johnson2019,Melnikov2018,Melnikov2020,Blank2023_SF-IRS,Zhang2023_THz-SFE} have been demonstrated and predicted that do not merely involve a direct coupling of light to the Raman-active phonon or magnon. Instead, the incoming photon first coherently excites an infrared (IR)-active phonon, which then scatters by the Raman-active phonon or magnon, and afterwards re-emits a photon, replacing the three-particle coupling term in the Raman tensor with phonon-phonon or spin-phonon coupling. The IR-active phonon therefore acts as a transducer that can effectively enhance the Raman scattering efficiency. These processes are now known as difference- and sum-frequency ionic Raman scattering \cite{forst:2011,Juraschek2018}. In addition, hybrid mechanisms involving a coherently excited phonon and a photon have been predicted \cite{Khalsa2021}, however an experimental verification is yet missing.


In this study, we develop a comprehensive framework for phonon-enhanced Raman scattering processes by phonons and magnons. For the case of scattering by phonons, we investigate the case of noncentrosymmetric materials, in which Raman-active phonons are also IR active and induce an electric polarization. For the case of scattering by magnons, we consider antiferromagnetic materials, in which the spin precession of the magnon induces a magnetization. We show that the difference-frequency, sum-frequency, and displacive excitations lead to difference-frequency generation, sum-frequency generation, and rectification of electric polarizations and magnetizations in the material, as depicted in Fig.~\ref{fig:mechanisms}b. These constitute a new class of phonon-mediated electro-optic and magneto-optic phenomena, which we call electro-phononic and magneto-phononic effects.

The remainder of the manuscript is organized as follows: In Sec.~\ref{sec:concept}, we introduce how coherent three-particle scattering leads to rectification, difference- and sum-frequency generation. In Sec.~\ref{sec:phonons}, we demonstrate the generation of nonlinear electric polarizations through electro-phononic effects. In Sec.~\ref{sec:magnons}, we demonstrate the generation of nonlinear magnetizations through magneto-phononic effects. In Sec.~\ref{sec:discussion}, we discuss the results. Details of all derivations are provided in the Appendix.


\section{Electric polarizations and magnetizations from coherent three-particle scattering processes}\label{sec:concept}

We begin by introducing how coherent three-particle scattering leads to rectification, difference- and sum-frequency generation. The potential energy for the three-particle scattering process can generally be written as
\begin{equation}
V=-aA_1A_2A_c, \label{eq:generalcoupling}
\end{equation}
where $a$ is a coupling coefficient and $A_1$ and $A_2$ represent either the electric field components of light ($A_i\equiv E_i$) or the amplitudes of coherent IR-active phonons ($A_i\equiv Q_i$). $A_c$ represents either the amplitude of a Raman-active phonon ($A_c \equiv Q_c$) or the magnetization component of a magnon ($A_c \equiv M$). The driving force acting on $A_c$ is generically given by
\begin{equation}
  F=-\frac{\partial V}{\partial A_c} =aA_1A_2, \label{eq:generalforce}
\end{equation}
which contains the sum or difference of the natural frequencies of the exciting fields $A_1$ and $A_2$, as well as a rectification component, as we will show in the following.

To illustrate the concept, we assume that the time dependence of $A_1$ and $A_2$ can be described by a cosine-shaped oscillation wrapped inside a Gaussian envelope, 
\begin{equation}
A_i(t) = \mathrm{e}^{\frac{-t^2}{2\sigma_i^2}} \cos(\omega_it) \label{eq:generaltimedependence}
\end{equation}
where $\sigma_i$ is the linewidth of particle $i \in \{1,2\}$, for example given by the duration of the laser pulse or the lifetime of the IR-active phonon, 
and $\omega_i$ is its natural frequency, for example given by the center frequency of the laser pulse or the eigenfrequency of the IR-active phonon. In the frequency domain, the driving force $F(\omega)$ is proportional to the convolution of the fields, $A_1(\omega)\circledast A_2(\omega)$, yielding
\begin{equation}
F(\omega) \propto e^{ -\frac{\sigma^2}{2}\left(\omega\pm(\omega_1-\omega_2)\right)^2 }  + e^{ -\frac{\sigma^2}{2}\left(\omega\pm(\omega_1+\omega_2)\right)^2 },\\
\end{equation}
where $\sigma^2 = \sigma_1^2\sigma_2^2/(\sigma_1^2+\sigma_2^2)$.
The driving force therefore exhibits peaks at the difference- and sum-frequency components, $|\omega_1-\omega_2|$ and $\omega_1+\omega_2$, which lead to difference-frequency generation (DFG) and sum-frequency generation (SFG). Furthermore, the driving force contains a quasistatic (DC) component, $F(\omega=0)$, given by 
\begin{equation}
F(\omega=0) \propto e^{ -\frac{\sigma^2}{2}\left(\omega_1-\omega_2\right)^2 } + e^{ -\frac{\sigma^2}{2}\left(\omega_1+\omega_2\right)^2 }.
\end{equation}
This quasistatic component is a displacive force, meaning $\langle A_1(t) A_2(t) \rangle \neq 0$. It leads to phononic rectification, inducing a quasistatic distortion of the crystal lattice \cite{forst:2011,subedi:2014}, as well as magnonic rectification, inducing a quasistatic spin canting \cite{kahana2023lightinduced}.

Table~\ref{tab:orderofexcitations} illustrates the scattering processes associated with three distinct excitation mechanisms for both phonons and magnons, in which the driving force acting on $A_c$ consists of two photons ($\nu-\nu$), two phonons ($ph-ph$), or one photon and one phonon ($\nu-ph$). The spectrum of each excitation has discernible peaks associated with DFG and SFG. A coherent oscillatory excitation of the $A_c$ mode occurs when its natural frequency is close to the peaks in the spectrum of the driving force. When the natural frequencies of the exciting particles are the same, $\omega_1 = \omega_2$, the observed peaks associated with DFG and SFG can be interpreted as rectification and second-harmonic generation (SHG), respectively.

\begin{table*}[t]
\centering
\def\arraystretch{1.3}
\caption{
\textbf{Three-particle scattering processes for coherent phonons and magnons.} Difference-frequency excitation mechanisms correspond to Raman scattering, whereas sum-frequency excitation mechanisms correspond to two-particle absorption processes. Processes involving two photons as exciting particles ($\nu-\nu$) for the Raman-active phonons or magnons are based on impulsive stimulated Raman scattering (ISRS) \cite{desilvestri:1985,merlin:1997,Kalashnikova2007,Kalashnikova2008,Tzschaschel2017,Afanasiev2021_anisotropy,Blank2023} and THz sum-frequency excitation (THz-SFE) \cite{maehrlein:2017,Juraschek2018,Knighton2019,Johnson2019,Giorgianni2022,Zhang2023_THz-SFE,Kusaba2024}. Processes involving two coherent IR-active phonons ($ph-ph$) as exciting particles are based on ionic Raman scattering (IRS) \cite{Wallis1971,forst:2011,subedi:2014,Neugebauer2021,Henstridge2022,nova:2017,Juraschek2021,Afanasiev2021,kahana2023lightinduced} and sum-frequency ionic Raman scattering (SF-IRS) \cite{Juraschek2018,Melnikov2018,Melnikov2020,Blank2023_SF-IRS}. Processes involving one photon and one phonon as exciting particles ($\nu-ph$) are based on infrared resonant Raman scattering (IRRS) and sum-frequency infrared resonant Raman scattering (SF-IRRS) \cite{Khalsa2021}. $\omega_i$ denotes photon frequencies, whereas $\Omega_i$ denotes IR-active phonon frequencies, and $i\in\{1,2\}$. $\Omega_c$ denotes the frequency of the Raman-active phonon or magnon.
}
\begin{tabular}{l K{2.375cm} K{2.375cm} K{2.375cm} K{2.375cm} K{2.375cm} K{2.375cm}}
\hline\hline
 & ISRS & THz-SFE & IRS & SF-IRS & IRRS & SF-IRRS\\
\hline
Type of process & Raman scattering & Two-particle absorption & Raman scattering & Two-particle absorption & Raman scattering & Two-particle absorption\\
Exciting particles & $\nu-\nu$ & $\nu-\nu$ & $ph-ph$  & $ph-ph$ & $\nu-ph$  & $\nu-ph$ \\
Center frequency & $\omega_i > \Omega_c$ & $\omega_i = \Omega_c/2$ & $\Omega_i > \Omega_c$ & $\Omega_i = \Omega_c/2$ & $\omega_i,\Omega_i > \Omega_c$  & $\omega_i,\Omega_i = \Omega_c/2$\\
Frequency mixing & Difference & Sum & Difference & Sum &  Difference & Sum \\
\hline
Phonon scattering & & & & & \\
\begin{minipage}{2.5cm}
\includegraphics[scale=0.1]{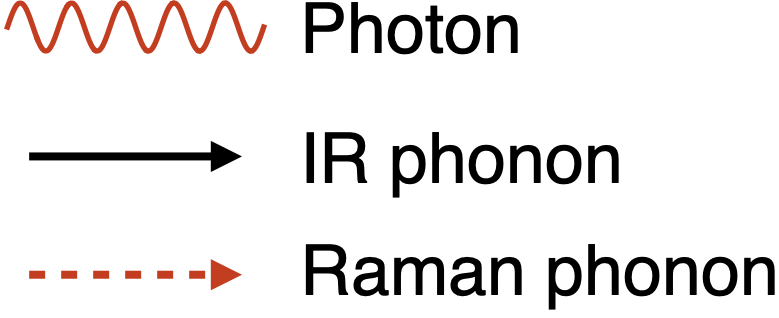}
\end{minipage}  
& 
\begin{minipage}{\linewidth}
\includegraphics[scale=0.08]{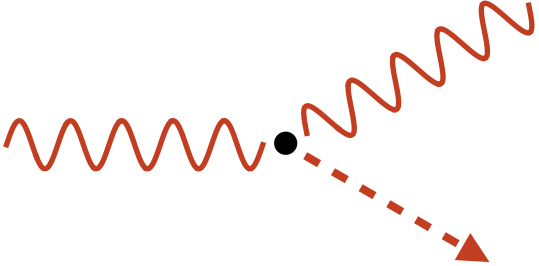}
\end{minipage} 
&
\begin{minipage}{\linewidth}
\includegraphics[scale=0.08]{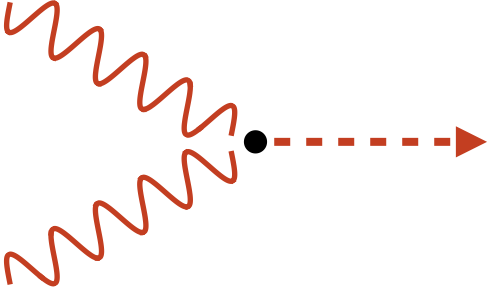} 
\end{minipage} 
&
\begin{minipage}{\linewidth}
\includegraphics[scale=0.08]{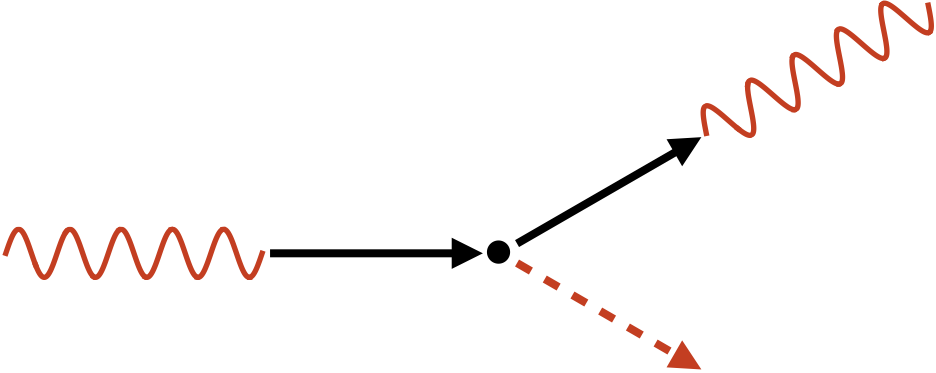}
\end{minipage} 
&
\begin{minipage}{\linewidth}
\includegraphics[scale=0.08]{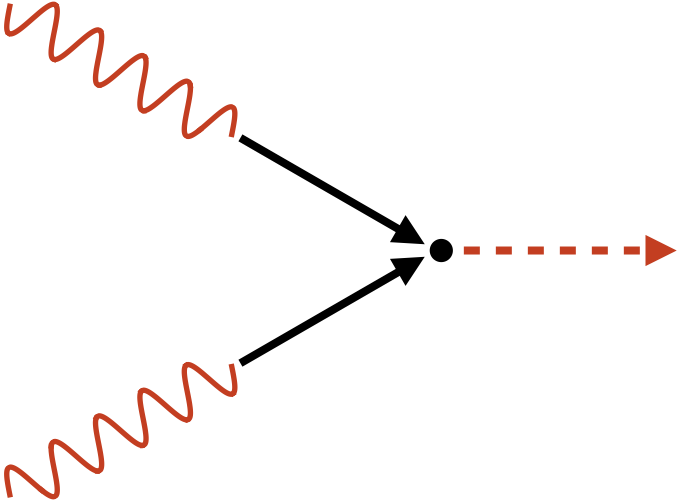}
\end{minipage} 
& 
\begin{minipage}{\linewidth}
\includegraphics[scale=0.08]{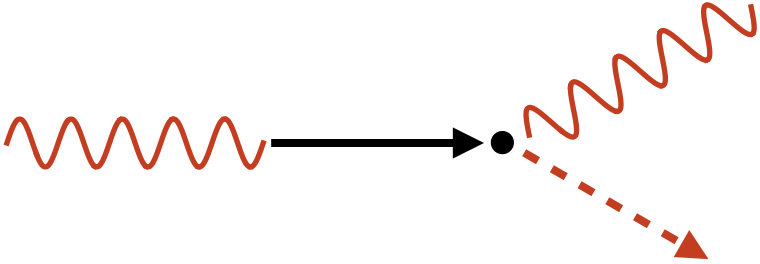}
\end{minipage} 
& 
\begin{minipage}{\linewidth}
\includegraphics[scale=0.08]{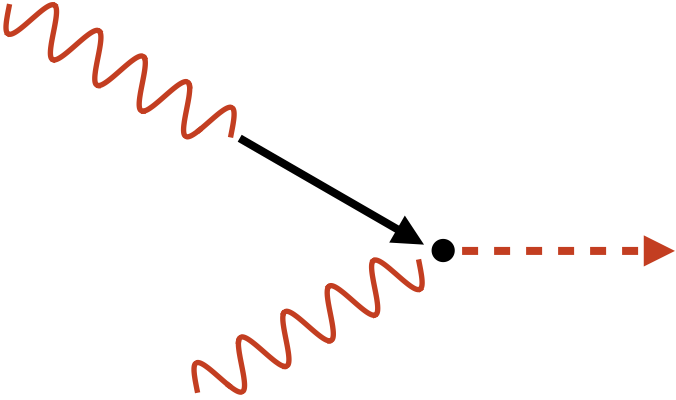}
\end{minipage}
\\ 
Force on phonon & $E^2$ & $E^2$ & $Q^2$ & $Q^2$ & $EQ$ & $EQ$\\
Example studies & \cite{desilvestri:1985,merlin:1997} & \cite{maehrlein:2017,Juraschek2018,Johnson2019,Knighton2019,Zhang2023_THz-SFE} & \cite{Wallis1971,forst:2011,subedi:2014,Mankowsky:2015} & \cite{Juraschek2018,Melnikov2018,Melnikov2020,Blank2023_SF-IRS} & \cite{Khalsa2021} &  \cite{Khalsa2021} \\
\hline
Magnon scattering & & & & & \\
\begin{minipage}{2.5cm}
\includegraphics[scale=0.1]{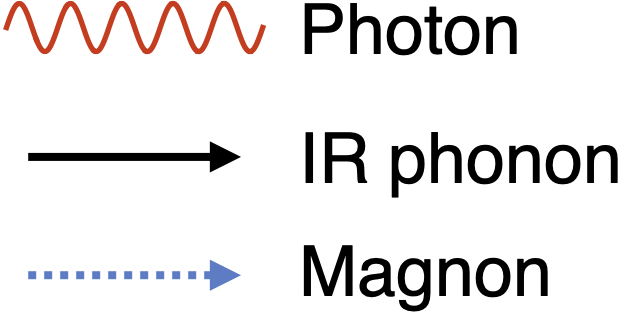}
\end{minipage} 
& 
\begin{minipage}{\linewidth}
\includegraphics[scale=0.08]{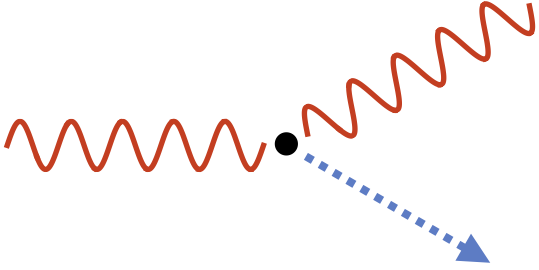}
\end{minipage} 
&
\begin{minipage}{\linewidth}
\includegraphics[scale=0.08]{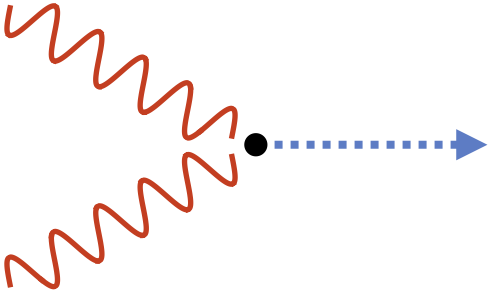} 
\end{minipage} 
&
\begin{minipage}{\linewidth}
\includegraphics[scale=0.08]{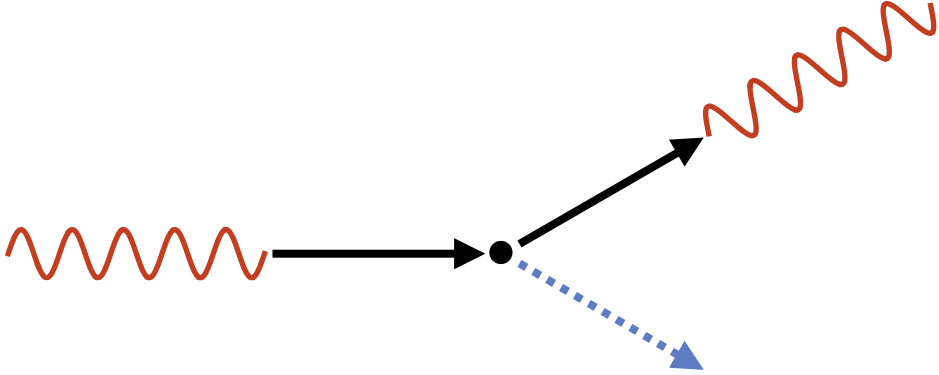}
\end{minipage} 
&
\begin{minipage}{\linewidth}
\includegraphics[scale=0.08]{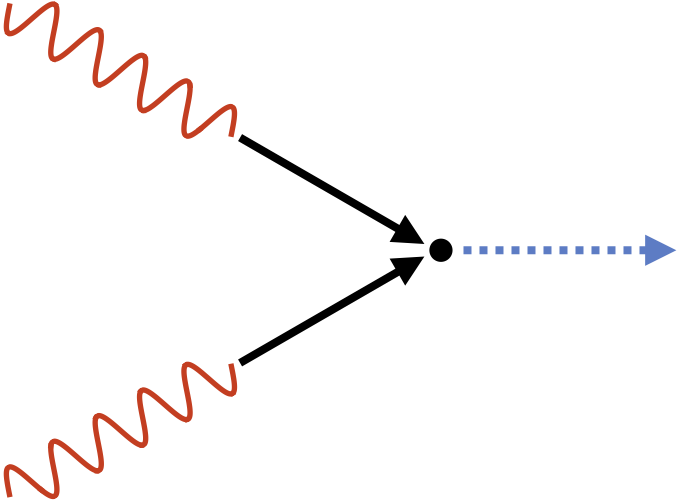}
\end{minipage} 
& 
\begin{minipage}{\linewidth}
\includegraphics[scale=0.08]{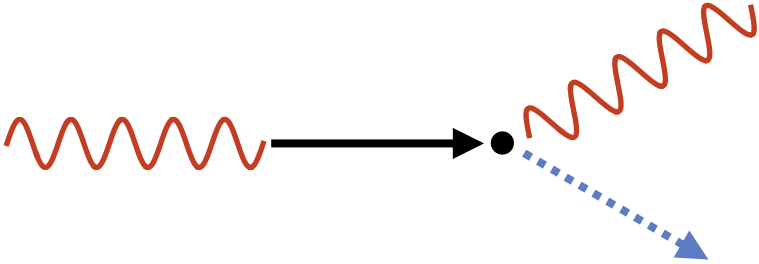}
\end{minipage} 
& 
\begin{minipage}{\linewidth}
\includegraphics[scale=0.08]{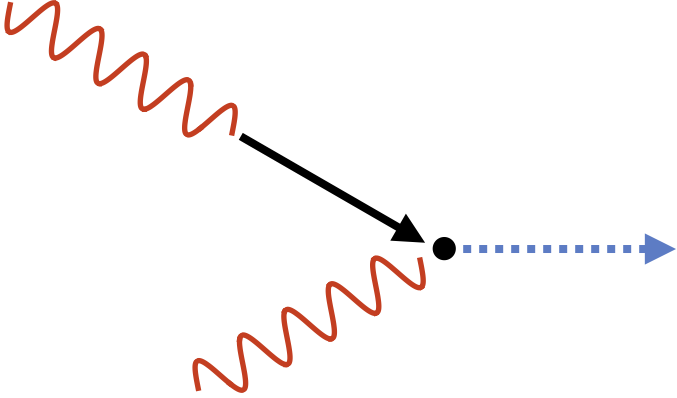}
\end{minipage}
\\ 
Force on magnon & $\mathbf{E}\times\mathbf{E}^*$ & $\mathbf{E}\times\mathbf{E}^*$  & $\mathbf{Q}\times\mathbf{Q}^*$ & $\mathbf{Q}\times\mathbf{Q}^*$  & $\mathbf{E}\times\mathbf{Q}^*$  & $\mathbf{E}\times\mathbf{Q}^*$ \\
Example studies & \cite{Kalashnikova2007,Kalashnikova2008,Tzschaschel2017} & \cite{Juraschek2021} & \cite{nova:2017,Juraschek2020_3,Juraschek2021,kahana2023lightinduced} & \cite{Juraschek2021} & -- & -- \\
\hline\hline
\end{tabular}
\label{tab:orderofexcitations}
\end{table*}
Eqs.~\eqref{eq:generalcoupling} and \eqref{eq:generalforce} are directly applicable when $A_c$ corresponds to a Raman-active phonon. For magnons, we will see that the coupling coefficient $a$ is an antisymmetric tensor that leads to an expression given by $V=a(A_1\dot{A_2}-\dot{A_1}A_2)A_{c}$, corresponding to a circularly polarized superposition of the exciting fields. As before, the spectrum of the driving force $F=A_1\dot{A_2}-\dot{A_1}A_2$ will have peaks at the difference- and sum-frequency components, $\omega_1\pm\omega_2$, associated with DFG and SFG, as shown in Table~\ref{tab:orderofexcitations}. However, if both particles have the same linewidth and natural frequency, $\sigma_1=\sigma_2$ and $\omega_1=\omega_2$, there will be only rectification, but no SHG.


\section{Electro-phononic effects}\label{sec:phonons}

We now develop the theoretical formalism for the electro-phononic effects from the case of scattering by Raman-active phonons, $A_c\equiv Q_c$. We derive the electric susceptibilities and the analytical time-dependent solutions for the second-order nonlinear polarizations. The investigated phenomena can be described by a second-order response of the phonon modes in the system to the electric field of the laser pulse. 
We will look at noncentrosymmetric systems, in which the Raman-active phonons are at the same time also infrared active and possess a zeroth-order electric dipole moment,
\begin{equation}\label{eq:dipolemoment}
    \mathbf{p}_c^{(0)} = \mathbf{Z}_c Q_c,
\end{equation}
where $\mathbf{Z}_c$ is the mode effective charge vector, given in units of $e$/$\sqrt{u}$, where $e$ is the elementary charge. The electric polarization per unit cell induced by the phonon mode is then given by $\mathbf{P}_c = \mathbf{p}_c^{(0)}/V_c$, where $V_c$ is the unit-cell volume. The coherent time evolution of a phonon mode can be generally described by a phenomenological oscillator model \cite{subedi:2014,fechner:2016,Juraschek2018} derived from the damped Euler-Lagrange equations
\begin{equation}
    \frac{\text{d}}{\text{d}t}\frac{\partial \mathcal{L}}{\partial \dot{Q}} - \frac{\partial \mathcal{L}}{\partial Q} = -\frac{\partial G}{\partial \dot{Q}}. \label{eq:phononeom}
\end{equation}
Here, $\mathcal{L}=T - V$ is the phonon Lagrangian containing the phonon kinetic energy, $T=\dot{Q}^2/2$, and the total phonon-dependent potential, $V$. $G=\kappa T$ is the Rayleigh dissipation function, where $\kappa$ is the phonon linewidth. 


\subsection{Two-photon excitation}

In two-photon excitations, two frequency components contained in the broad spectrum of an ultrashort laser pulse interact simultaneously with a Raman-active phonon and excite it coherently when their difference or sum matches the frequency of the phonon mode \cite{merlin:1997,maehrlein:2017}, as shown in Fig.~\ref{fig:mechanisms}. These correspond to impulsive stimulated Raman scattering (ISRS) \cite{desilvestri:1985,merlin:1997} and terahertz sum-frequency excitation (THz-SFE) \cite{maehrlein:2017,Juraschek2018}, respectively, as listed in  Table~\ref{tab:orderofexcitations}. Here, virtual or real electronic states serve as the intermediate state in the scattering process. When the photon energy is larger than the band gap of the material, this will in addition create a nonequilibrium electronic configuration, which leads to a displacive excitation of coherent phonons (DECP) \cite{zeiger:1992,Giorgianni2022}, see Fig.~\ref{fig:mechanisms}. In the following, we will consider the photon energy to be well below the band gap.

The coupling in Eq.~\eqref{eq:generalcoupling} contains the electric field components of the ultrashort pulse, $A_{i}\equiv E_{i}$. A minimal model can be written as
\begin{equation}
    V = \frac{\Omega_c^2}{2}Q_c^2 - R_{ij}E_iE_jQ_c,
\end{equation}
where $Q_c$ is the Raman-active phonon amplitude and $\Omega_c$ is its eigenfrequency. $Q_c$ is in units of \AA$\sqrt{u}$, where $u$ is the atomic mass unit. The light-matter coupling strength is given by the Raman tensor, $R_{ij}$, which describes the change in electronic polarizability by the Raman-active phonon, $R_{ij}=\partial \varepsilon_{ij}/\partial Q_c$. The indices $i$ and $j$ denote spatial coordinates and we use the Einstein sum convention for their summation. To investigate the time evolution of the Raman-active phonon mode, we solve its equation of motion, which according to Eq.~\eqref{eq:phononeom} becomes
\begin{equation}
\ddot{Q}_c + \kappa_c \dot{Q}_c + \Omega_c^2 Q_c = R_{ij}E_i(t)E_j(t).
\label{eq:1}
\end{equation}
Here, $R_{ij}E_i(t)E_j(t)$ acts as the driving force for the phonon. We choose our coordinate system so that $R_{ij}E_i(t)E_j(t)\equiv R_{zz}E^2(t)$ and we model the electric field component of the laser pulse according to
\begin{equation}
    E_i(t) = E_0 \exp\left(-{\frac{t^2}{2 \sigma^2}}\right)  \cos(\omega_i t + \phi),
\label{eq:pulse}
\end{equation} 
where $E_0$ is the peak electric field, $\omega_i$ is the center frequency, $\sigma = \tau/\sqrt{8\ln 2}$, where $\tau$ is the full width at half maximum pulse duration, and $\phi$ is the carrier envelope phase. We assume that the laser pulse has no spectral overlap with the Raman-active phonon mode, so that an excitation through the zeroth-order electric dipole moment, Eq.~\eqref{eq:dipolemoment}, can be excluded.

If we Fourier transform Eq.~\eqref{eq:1} to the frequency domain, we obtain the amplitude of the Raman-active phonon mode to second order in the electric field, 
\begin{equation}
Q_c^{(2)}(\omega) =  \frac{R_{ij}}{\Delta_c(\omega)} \left( E_i(\omega)\circledast E_j(\omega) \right),
\end{equation}
where $\Delta_c(\omega)=\Omega_c^2-\omega^2+i\kappa_c\omega$. We can further obtain an analytical expression for the phonon amplitude in the time domain, given by Eq.~\eqref{eq:solQc_2_2photon_analytic} in the Appendix. Since in a noncentrosymmetric material, the Raman-active phonon mode carries an electric dipole moment given by Eq.~\eqref{eq:dipolemoment}, the two-photon excitation therefore induces a second-order nonlinear electric polarization that can be written as
\begin{align}
P_{c,i}^{(2)} = 
\frac{\varepsilon_0}{\sqrt{2\pi}}\int\limits_{-\infty}^{\infty} \chi_{e,ijk}^{(2)}(\omega,\omega')E_j(\omega-\omega')E_k(\omega')d\omega',
\label{eq:Pe1}
\end{align}
where $\chi_{e,ijk}^{(2)}$ is a second-order nonlinear electric susceptibility containing the light-matter coupling of the phonon,
\begin{equation}
\chi_{e,ijk}^{(2)}(\omega,\omega') = \frac{Z_{c,i}\left({R_{jk}}+{R_{kj}}\right)}{2\varepsilon_0V_c\Delta_c(\omega)}.
\label{eq:Xe1}
\end{equation}
The norm of $\chi_{c,ijk}^{(2)}(\omega)$ has a peak at the eigenfrequency of the phonon mode. Note that the Raman tensors $R_{ij}$ further have implicit electronic resonances at and above the band-gap energy. We assume here that the band gap is much larger than the excitation frequencies, away from these electronic resonances. 

In the remainder of the manuscript, we will investigate the characteristics of each of the susceptibilities utilizing the average
\begin{equation}
 \chi_{e,ijk}^{(2)}(\omega) \equiv \frac{1}{\Delta \omega} \int_{\Delta \omega} \chi_{e,ijk}^{(2)}(\omega,\omega') d\omega'
\label{eq:Xe_av}
\end{equation}
within a frequency range $\Delta \omega$, encompassing all resonant frequencies inherent to the system.


\subsection{Two-phonon excitation}

Instead of interacting with the Raman-active phonon directly, two frequency components from an ultrashort laser pulse can couple to IR-active phonons, which serve as the intermediate state for the scattering process. A coherent excitation of the Raman-active phonon is achieved if its frequency matches the difference or sum of the frequencies of the two IR-active phonons, corresponding to ionic Raman scattering (IRS) \cite{Wallis1971,forst:2011,subedi:2014} and sum-frequency ionic Raman scattering (SF-IRS) \cite{Juraschek2018}, respectively, see Table~\ref{tab:orderofexcitations}. In addition, the two-phonon process always leads to a displacive excitation as long as the IR-active phonons are ringing, even when the photon energy is below the band gap and no nonequilibrium electronic configuration is created \cite{subedi:2014,mankowsky:2014}. The IR-active phonons therefore act as transducers, in which the coupling to the Raman-active phonon is provided by anharmonicities of the interatomic potential energy surface, instead of electron-phonon coupling as in the two-photon excitation process. The coupling in Eq.~\eqref{eq:generalcoupling} here contains the amplitudes of a driven IR-active phonon mode, $A_i\equiv Q_i$ and a minimal model can accordingly be written as 
\begin{align}
V = &~ \frac{\Omega_c^2}{2}Q_c^2+\frac{\Omega_1^2}{2}Q_1^2+\frac{\Omega_2^2}{2}Q_2^2 \nonumber\\
& - Q_1 Z_{1,i}E_i - Q_2 Z_{2,i}E_i -c Q_1 Q_2 Q_c,
\end{align}
where $\Omega_{1/2}$ are the eigenfrequencies and $Q_{1/2}$ the amplitudes of the driven IR-active phonon modes. These can either correspond to phonon modes with different symmetries and frequencies ($\Omega_1 \neq\Omega_2$), or to the same phonon mode ($\Omega_1 =\Omega_2 \equiv \Omega_0$). The nonlinear phonon coupling coefficient $c$ is given in units of meV/(\AA{}$\sqrt{u}$)$^3$. The equations of motion derived from Eq.~\eqref{eq:phononeom} yield 
\begin{subequations}
\begin{align}
\ddot{Q}_c + \kappa_c \dot{Q}_c + \Omega_c^2 Q_c  = &~ cQ_1 Q_2,
\label{eq:2a}
\\
\ddot{Q}_1 + \kappa_1 \dot{Q}_1 + \Omega_1^2 Q_1  = &~ Z_{1,i}E_i(t) +cQ_c Q_2,
\label{eq:2b}
\\
\ddot{Q}_2 + \kappa_2 \dot{Q}_2 + \Omega_2^2 Q_2  = &~ Z_{2,i}E_i(t) +cQ_c Q_1.
\label{eq:2c}
\end{align}
\end{subequations}
The IR-active phonons are driven directly through their zeroth-order electric dipole moments, Eq.~\eqref{eq:dipolemoment}, with mode effective charges $Z_{1,i}$ and $Z_{2,i}$. Again, we assume the spectral overlap between the ultrashort laser pulse and the Raman-active phonon to be negligible, which leaves $cQ_1 Q_2$ as its sole driving force. Transforming Eqs.~\eqref{eq:2b} and \eqref{eq:2c} to the frequency domain, we obtain the linear response of the IR-active phonons to the electric field, whereas transforming \eqref{eq:2a} yields the second-order response of the Raman-active phonon mode, 
\begin{subequations}
\begin{align}
Q_{1}^{(1)}(\omega)&=\frac{Z_{1,i}}{\Delta_{1}(\omega)}E_i(\omega),~ Q_{2}^{(1)}(\omega) =\frac{Z_{2,i}}{\Delta_{2}(\omega)}E_i(\omega) \label{eq:3a} \\
Q_c^{(2)}(\omega) &= \frac{c}{\Delta_c(\omega)} \left( Q_1^{(1)}(\omega) \circledast Q_2^{(1)}(\omega)\right) \nonumber\\
&= \frac{cZ_{1,i}Z_{2,j}}{\Delta_c(\omega)} \left(\frac{E_i(\omega)}{\Delta_1(\omega)} \circledast \frac{E_j(\omega)}{\Delta_2(\omega)} \right) \label{eq:3b}
\end{align}
\end{subequations}
A Fourier transformation back to the time domain yields a semi-analytical expression for the phonon amplitude, given by Eq.~\eqref{eq:time_Qc2} in the Appendix.

As for the two-photon excitation mechanism, we can calculate the electric polarization induced by the Raman-active phonon mode in a noncentrosymmetric material given by Eq.~\eqref{eq:Pe1}, with the second-order nonlinear electric susceptibility given by the nonlinear phonon and light-matter couplings, 
\begin{equation}
\begin{split}
\chi_{e,ijk}^{(2)}(\omega,\omega') = &~ \frac{cZ_{c,i}}{2\varepsilon_0V_c\Delta_c(\omega)}\left[\frac{Z_{1,j}Z_{2,k}}{\Delta_1(\omega-\omega')\Delta_2(\omega')}\right. \\
& + \left. \frac{Z_{1,k}Z_{2,j}}{\Delta_1(\omega')\Delta_2(\omega-\omega')} \right].
\end{split}
\end{equation}


\subsection{One-photon-one-phonon excitation}

One-photon-one-phonon excitations combine features from the two-photon and two-phonon excitation mechanisms by inducing changes in the electronic and ionic contributions to the polarizability. To excite a Raman-active phonon coherently, its frequency needs to match the difference or sum of the frequencies of a photon from the laser pulse and of an IR-active phonon driven by it. These mechanisms have been proposed as infrared resonant Raman scattering (IRRS) and sum-frequency infrared resonant Raman scattering (SF-IRRS) recently \cite{Khalsa2021}, see Table~\ref{tab:orderofexcitations}, however no experimental measurement has yet been demonstrated. The coupling in Eq.~\eqref{eq:generalcoupling} here contains the electric field component of light $A_1\equiv E_i$ and the amplitude of the driven IR-active phonon mode, $A_2\equiv Q_2$. A minimal model can be written as
\begin{equation}
V=\frac{\Omega_c^2}{2}Q_c^2+\frac{\Omega_2^2}{2}Q_2^2-Q_2 Z_{2,i}E_i-b_iE_iQ_cQ_2, \\
\end{equation}
where $b_i$ is a vector describing the hybrid nonlinear photon-phonon coupling. The light-matter coupling terms in the above equation can be summarized as $\textbf{p}_2^{(0)}(Q_c) = (\textbf{Z}_2 + \textbf{b}Q_c)Q_2$, showing that the components $b_i$ arise from modifications of the mode effective charge of the IR-active phonon by the Raman-active phonon. The equations of motion derived from Eq.~\eqref{eq:phononeom} yield
\begin{subequations}
\begin{align}
\ddot{Q}_c + \kappa_c \dot{Q}_c + \Omega_c^2 Q_c  &= b_iE_i(t)Q_2,
\label{eq:4a}\\
\ddot{Q}_2 + \kappa_2 \dot{Q}_2 + \Omega_2^2 Q_2  &= Z_{2,i}E_i(t)+b_iE_i(t) Q_c.
\label{eq:4b}
\end{align}
\end{subequations}
As in the two-phonon excitation process, the electric field component drives the IR-active phonon mode directly via the electric dipole coupling. Assuming no spectral overlap between the ultrashort laser pulse and the Raman-active phonon, the force acting on it is given by $b_iE_iQ_2$. The response of $Q_{2}$ is predominantly linear in the electric field and given by the same expression as in Eq.~\eqref{eq:3a}. The Fourier transform of Eq.~\eqref{eq:4a} combined with the expression for $Q_{2}(\omega)$ yields the second-order nonlinear contribution of the amplitude of the Raman-active phonon in the frequency domain,
\begin{align}
Q_c^{(2)}(\omega) &= \frac{b_i}{\Delta_c(\omega)} \left( E_i(\omega) \circledast Q_2^{(1)}(\omega) \right) \nonumber\\
&= \frac{b_iZ_{2,j}}{\Delta_c(\omega)} \left( E_i(\omega)\circledast \frac{E_j(\omega)}{\Delta_2(\omega)} \right).
\end{align}
A Fourier transformation back to the time domain yields a semi-analytical expression for the phonon amplitude, given by Eq.~\eqref{eq:time_Qc2} in the Appendix. As in the previous sections, $Q_c^{(2)}(\omega)$ produces an electric polarization according to Eq.~\eqref{eq:Pe1}, where here the second-order nonlinear electric susceptibility contains the nonlinear photon-phonon coupling,
\begin{align}
\chi_{e,ijk}^{(2)}(\omega,\omega') = & 
\frac{Z_{c,i}}{2\varepsilon_0V_c\Delta_c(\omega)} \nonumber\\
& \times \left(\frac{b_jZ_{2,k}}{\Delta_2(\omega')} +\frac{b_kZ_{2,j}}{\Delta_2(\omega-\omega')}\right).
\end{align}
Note that the coefficient $b_i$ further has an implicit electronic resonance at and above the band-gap energy. We assume here that the band gap is much larger than the excitation frequencies, away from this resonance. 


\subsection{Examples of electro-phononic polarizations and susceptibilities}

\begin{figure*}[t]
    \centering


    
    
    \includegraphics[width=1\textwidth]{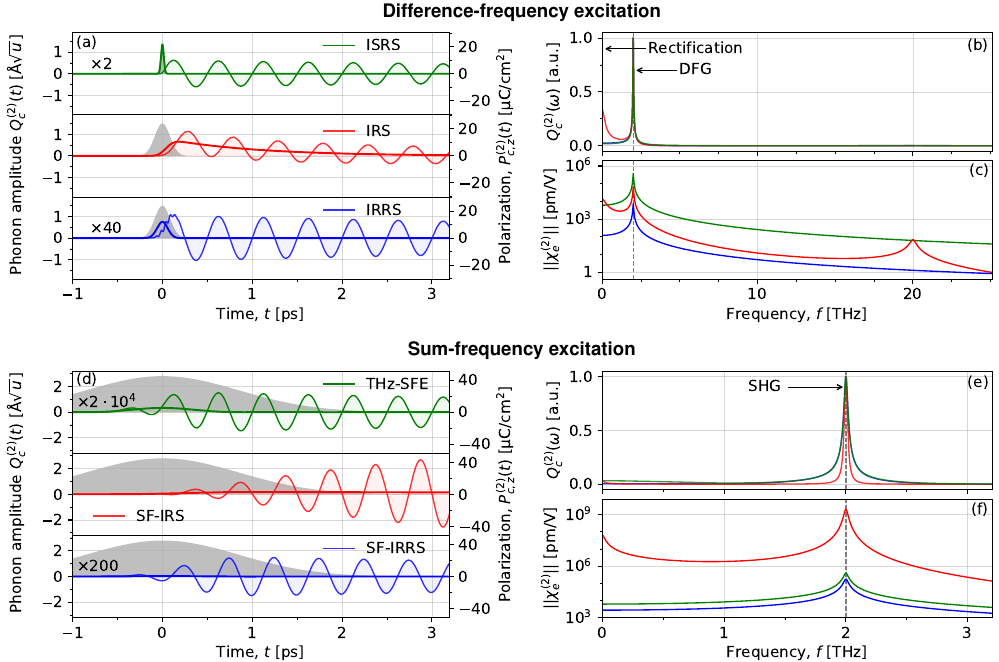}
    \caption{
    \textbf{Electro-phononic effects for the example of a 2~THz Raman-active phonon mode.}
    (a--c) Demonstration of phononic rectification and difference-frequency generation (DFG) arising from impulsive stimulated Raman scattering (ISRS), ionic Raman scattering (IRS), and infrared resonant Raman scattering (IRRS). For ISRS, we set the center wavelength of the pulse to 800~nm, the peak electric field to $E_0=20$~MV/cm, and the pulse duration to $\tau=50$~fs. For IRS and IRRS, we set the eigenfrequencies of the IR-active phonon modes and therefore the center frequencies of the mid-IR pulses to $\Omega_i/(2\pi)=\omega_i/(2\pi)=10$ THz, keeping the total pulse energy constant by setting $E_0=10$~MV/cm and $\tau=200$~fs. 
    (a) Time-dependent phonon amplitude and polarization for each of the three mechanisms. The pulse duration is indicated as grey shaded area. 
    (b) Normalized Fourier transforms of the time traces in (a). 
    (c) Norm of the averaged nonlinear electric susceptibility.
    (d--f) Demonstration of sum-frequency generation and specifically second-harmonic generation (SHG) arising from terahertz sum-frequency excitation (THz-SFE), sum-frequency ionic Raman scattering (SF-IRS), and sum-frequency infrared resonant Raman scattering (SF-IRRS). We take the frequencies of the exciting particles to be half the Raman-active phonon frequency, $\Omega_i/(2\pi)=\omega_i/(2\pi)=1$~THz. For each THz-SFE, SF-IRS, and SF-IRRS, we set $E_0=0.1$~MV/cm and $\tau=1$~ps.
    (d) Time-dependent phonon amplitude and polarization for each of the three mechanisms. 
    (e) Normalized Fourier transforms of the time traces in (d). 
    (f) Norm of the nonlinear averaged electric susceptibility.
    }
    \label{fig:main1}
\end{figure*}

We now evaluate the formalism developed in the previous sections for the three mechanisms, using typical values for the phonon-mode and light-matter interactions parameters, displayed in Table~\ref{tab:parameters1}. In Fig.~\ref{fig:main1}, we show the time evolutions of the electric polarizations arising from the Raman-active phonon amplitudes, $P_{c,i}(t)$, according to the three mechanisms. We use the example of a Raman-active phonon mode with an eigenfrequency of $\Omega_c/(2\pi)=2$~THz. We distinguish two cases: In Fig.~\ref{fig:main1}(a--c), we take the frequencies of the exciting particles to be larger than the Raman-active phonon frequency, $\omega_i, \Omega_{i} > \Omega_c$, in order to illustrate electro-phononic difference-frequency generation and rectification. In Fig.~\ref{fig:main1}(d--f), we take the frequencies of the exciting particles to be half of the Raman-active phonon frequency, $\omega_i, \Omega_{i} = \Omega_c/2$, in order to illustrate electro-phononic sum-frequency generation, specifically second-harmonic generation. The two-photon mechanisms are plotted in green, the two-phonon mechanisms in red, and the one-photon-one-phonon mechanisms in blue. We use the analytical solutions of the Raman-active phonon amplitude according to Eq.~\eqref{eq:time_Qc2} in the Appendix. The analytical solutions match remarkably well with numerical evaluations of the respective equations of motion, as shown in Fig.~\ref{fig:analyticalcomparisonphonons} in the Appendix. Only for large coupling strengths, a renormalization of the phonon frequencies creates a deviation from the numerical results. For the purpose of our study, all relevant features are captured by the analytical expressions however.


\begin{table}[t]
\def\arraystretch{1.3}
\centering
\caption{\textbf{Parameters for electro-phononic effects.} $R_{zz}$, $c$, and $b_z$ are the coupling strengths of the three-particle scattering processes, two-photon, two-phonon, and
one-photon-one-phonon, respectively. $Z_{\sigma,z}$, $\Omega_\sigma$, and $\kappa_\sigma$ is the mode effective charge, eigenfrequency and linewidth, respectively, of phonon mode $\sigma$, and $V_c$ is the volume of the unit cell.
}
\begin{tabular}{cccccc}
\hline\hline
$R_{zz}$ & $c$ & $b_z$ &  $Z_{\sigma,z}$ & $\kappa_\sigma$ & $V_c$ \\ 
100~$\varepsilon_0\frac{\text{\AA}^2}{\sqrt{u}}$ & 10~$\frac{\text{meV}}{(\text{\AA}\sqrt{u})^3}$ & $0.1$ $\frac{e}{\text{\AA}u}$ & 1~$\frac{e}{\sqrt{u}}$ & $0.1\frac{\Omega_\sigma}{2\pi}$ & 100~\AA$^3$\\
\hline\hline
\end{tabular}
\label{tab:parameters1}
\end{table}


Fig.~\ref{fig:main1}(a) shows the temporal evolution of the phonon amplitudes for the three difference-frequency processes. The excitation induced by the difference-frequency components of each of the mechanisms causes an impulsive response of the phonon amplitude and therefore polarization. Our analysis assumes that the excitation energy lies well below the band-gap energy. Consequently, for mechanisms directly involving the electric field component of light, such as impulsive stimulated Raman scattering (ISRS) and infrared resonant Raman scattering (IRRS), rectification occurs solely during the duration of the laser pulse, which here is less than one oscillation period of the Raman-active phonon. On the other hand, ionic Raman scattering (IRS) induces a prolonged rectified state as long as the driven phonon modes continue ringing. In Fig. \ref{fig:main1}(b), we depict the normalized Fourier transforms of the temporal traces illustrated in (a). Distinct and symmetrical DFG peaks emerge at the eigenfrequency of the coupled phonon mode at 2~THz. Additionally, a static component is observed at zero frequency for IRS, indicating the unidirectional force acting on the coupled phonon. For ISRS and IRRS, the static component is nonzero, but negligible. 
In Fig.~\ref{fig:main1}(c), we plot the norm of the average of the nonlinear electric susceptibilities given by Eq.~\eqref{eq:Xe_av}. Here, three peaks of rectification, DFG, and SHG can be distinguished for the IRS mechanism, whereas ISRS and IRRS show only a peak at the eigenfrequency of the Raman-active phonon.

Fig.~\ref{fig:main1}(d) shows the temporal evolution of the phonon amplitudes for the three sum-frequency, specifically second-harmonic generation, processes. Notably, the phonon amplitude and polarization show a gradual build-up of the, rather than an impulsive response, characteristic for sum-frequency excitation. For all three sum-frequency mechanisms, rectification is negligible compared to the oscillating part of the phonon amplitude and polarization. In Fig. \ref{fig:main1}(e), we present the normalized Fourier transforms of the temporal traces from (d). Symmetrical peaks can be seen at the eigenfrequency of the Raman-active phonon at 2~THz, here induced by the SFG/SHG components of the respective driving forces. These peaks are well visible in Fig.~\ref{fig:main1}(f), where we plot the norm of the averaged nonlinear electric susceptibilities. In the logarithmic scaling of the plot, a small rectification component of the SF-IRS mechanism can further be identified.

All calculations performed in Fig.~\ref{fig:main1} are continuously scalable to higher pulse fluences. In reality however, the amount of induced phonon amplitude and therefore polarization is limited by the stability of the crystal lattice. The Lindemann criterion \cite{Lindemann1910,sokolowski-tinten:2003} is a common estimator for the stability of the lattice in coherent phonon excitations and requires that the mean-squared atomic displacements remain below 10-20\%{} of the interatomic distance. The maximum amplitudes of $Q\approx 2$~\AA$\sqrt{u}$ shown here are typically below this limit \cite{Juraschek2018}. The relative strengths of the photon-photon, phonon-phonon, and photon-phonon processes are strongly material dependent and are evaluated here for parameters typical for perovskite oxides (Table~\ref{tab:parameters1}). The amplitudes of the IR-active phonon modes in the two-phonon and one-photon-one-phonon processes are shown in Fig.~\ref{fig:Qd1} in the Appendix. 


\subsection{2D nonlinear electric susceptibility}

\begin{figure}[h]
    \centering
    
    

    
    \includegraphics[width=1.\linewidth]{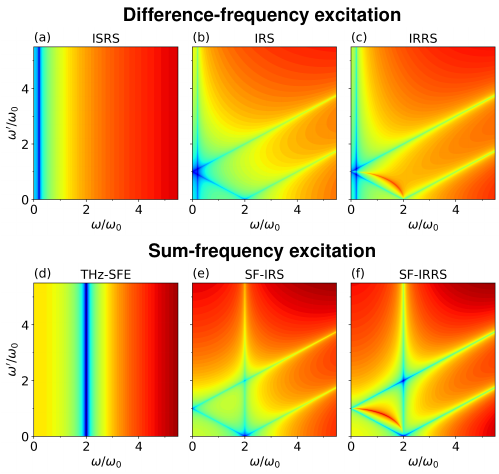}
    \caption{
    \textbf{2D nonlinear electric susceptibility.} We show the norm of the second-order nonlinear electric susceptibility, $||\chi_{e,zzz}(\omega,\omega')||$ on a logarithmic scale for the three difference-frequency (a--c) and three sum-frequency (d--f) excitation mechanisms, respectively. More intense blue indicates larger values, more intense red indicates smaller values, plotted in arbitrary units.
    }
    \label{fig:HM1}
\end{figure}

In Fig.~\ref{fig:HM1}, we present the norm of the electric susceptibility denoted as $||\chi_{e,zzz}(\omega,\omega')||$. Within Fig. \ref{fig:HM1}(a--c), the frequencies attributed to the exciting particles exceed the Raman-active phonon frequency, $\Omega_i > \Omega_c$. This scenario leads to phononic difference-frequency generation and rectification. The Raman-active phonon mode is characterized by an eigenfrequency of $\Omega_c/(2\pi) = 2$~THz, while the wavelength is set to 800~nm for the two-photon process and $\omega_i/(2\pi)=\Omega_i/(2\pi)=10$~THz for two-phonon and one-photon-one-phonon processes. (a), (b), and (c) correspond to ISRS, IRS, and IRRS, respectively. In Fig.~\ref{fig:HM1}(d--f), the frequencies associated with the exciting particles are half that of the Raman-active phonon frequency, indicated as $\omega_i = \Omega_i = \Omega_c/2$, resulting in electro-phononic sum-frequency generation, specifically second-harmonic generation. With a Raman-active phonon  frequency of $\Omega_c/(2\pi) = 2$~THz, we set $\omega_i/(2\pi) = \Omega_i/(2\pi)=1$~THz for all three mechanisms. (d), (e), and (f) correspond to THz-SFE, SF-IRS, and SF-IRRS, respectively. 


\section{Magneto-phononic effects}\label{sec:magnons}

We now develop an analog formalism for magneto-phononic rectification, and difference- and sum-frequency generation, based on two-photon, two-phonon, and one-photon-one-phonon excitations of magnons. We derive the magneto-electric susceptibilities and the analytical time-dependent solutions for the second-order nonlinear magnetizations. While the dynamics of the Raman-active phonons in the sections above is described by a phenomenological oscillator model, the spin precession dynamics of magnon modes can be described by the Landau-Lifshitz-Gilbert equations.

We investigate the example of a simple easy-plane Heisenberg antiferromagnet, describing a variety of prototypical antiferromagnetic systems such as NiO, for which the spin Hamiltonian can be written as

\begin{align}
\mathcal{H} =&~ J \textbf{S}_1 \cdot \textbf{S}_2 + D_x\left(S_{1,x}^2+S_{2,x}^2 \right)+ D_y\left(S_{1,y}^2+S_{2,y}^2 \right) \nonumber \\
& - \gamma_{el}\left(\textbf{S}_1+\textbf{S}_2 \right)\cdot \textbf{B}(t),
\end{align}
where $J>0$ is the exchange constant of the interaction between spins $\textbf{S}_1$ and $\textbf{S}_2$, $D_x>0$ and $D_y>0$ are uniaxial anisotropy constants, and $\gamma_{el}$ is the electron gyromagnetic ratio. This system hosts two magnons with eigenfrequencies $\hbar\Omega_l=\sqrt{D_x(J+D_y)}$ and $\hbar\Omega_h=\sqrt{D_y(J+D_x)}$, where $\Omega_l<\Omega_h$ for $D_x<D_y$. At equilibrium and in the absence of external fields, the spins $\textbf{S}_1$ and $\textbf{S}_2$ are aligned antiparallel along the $z$-axis. The interaction with an external magnetic field aligned along the $x$-axis, $\textbf{B}=B\hat{\textbf{x}}$, breaks the relationship $\textbf{S}_1 = - \textbf{S}_2$ between the spins and induces transient spin components along the $x$- and $y$-axes, which satisfy $S_{1,x}=S_{2,x}$, $S_{1,y}=-S_{2,y}$, and $S_{1,z}=-S_{2,z}$. This, in turn, enables the emergence of a transient magnetization along the $x$-axis, $M_x$. An external magnetic field pointing along the $x$-axis couples to the high-frequency magnon. 

The Zeeman-like term $(\textbf{S}_1+\textbf{S}_2)\cdot \textbf{B}(t)$ contains the three-particle scattering term in Eq.~\eqref{eq:generalcoupling}, where  $A_c\equiv \mathbf{M}=\frac{\hbar\gamma_{el}}{V_c}(\textbf{S}_1+\textbf{S}_2)$. $\textbf{B}$ is an effective magnetic field produced by the exciting particles, $A_1$ and $A_2$, acting as the driving force of the magnon according to Eq.~\eqref{eq:generalforce}. To study spin dynamics, we use the Landau-Lifshitz-Gilbert equations
\begin{equation}
\label{eq:LLG}
\frac{d \textbf{S}_\sigma}{dt} =  \frac{\gamma_{{el}}}{1+\kappa_{{el}}^2}\left[ \textbf{S}_\sigma \times \textbf{B}_\sigma^{\text{eff}} -\frac{\kappa_{{el}}}{|\textbf{S}_\sigma|} \textbf{S}_\sigma \times \left( \textbf{S}_\sigma \times \textbf{B}_\sigma^{\text{eff}}  \right) \right], \\
\end{equation}
where $\sigma \in \{1,2\}$, $\kappa_{el}$ is a phenomenological damping parameter, and $\textbf{B}_\sigma^{\text{eff}} = -(\hbar\gamma_{el})^{-1} \partial \mathcal{H}/\partial \textbf{S}_\sigma$. We solve the Landau-Lifshitz-Gilbert equations perturbatively, assuming that the effective magnetic field $\textbf{B}$ is induced by the electric field component of the laser pulse in quadratic order, $B \sim \mathcal{O}(E^2)$, which yields

\begin{equation}
M_x(\omega) = \frac{2\hbar\gamma_{el}^2 S}{V_c\left(1+\kappa_{el}^2\right)} \frac{i\kappa_{el}\omega+2SD_y/\hbar}{\Delta_m(\omega)}B(\omega), 
\label{eq:M_general_frequency}
\end{equation}
where $|\textbf{S}_\sigma|=S$ and $\Delta_m = \Omega_m^2-\omega^2+i\kappa_m\omega$. $\Omega_m=\frac{2}{\hbar}\sqrt{\frac{(J+D_x)D_y}{1+\kappa_{el}^2}}$ is the damping-renormalized magnon frequency, and $\kappa_m=\frac{2\kappa_{el}}{\hbar\left(1+\kappa_{el}^2\right)}\left(J+D_x+D_y\right)$ is the magnon linewidth. For a detailed derivation of this expression, see the Appendix. The effective magnetic field can arise as a result of three types of excitations, two-photon, two-phonon, and one-photon-one-phonon.


\subsection{Two-photon excitation}

Two-photon excitation processes for magnons are analog to those for phonons, with ISRS representing a difference-frequency process \cite{Kalashnikova2007,Kalashnikova2008}, THz-SFE a sum-frequency process \cite{maehrlein:2017,Juraschek2021}, and displacive excitation of coherent magnons (DECM) a displacive process \cite{Kalashnikova2008}. We list the scattering processes in Table~\ref{tab:orderofexcitations}. In contrast to the scattering by phonons, scattering by magnons additionally requires the transfer of angular momentum, which is achieved with circularly polarized light. The coupling in Eq.~\eqref{eq:generalcoupling} contains the electric field component of light, $A_i\equiv E_i$ and the coupling coefficient $a$ is an antisymmetric tensor. The coupling therefore requires a circularly polarized electric field of the laser pulse, which produces an effective magnetic field that drives the spin precession in the Landau-Lifshitz-Gilbert equations, Eq.~\eqref{eq:LLG}. Phenomenologically, the effective magnetic field can be written as 
\begin{equation}
    \textbf{B}(t) = \tilde{\alpha}
    \textbf{E}(t) \times \dot{\textbf{E}}(t),
\label{eq:B1}
\end{equation}
which is commonly known as the inverse Faraday effect \cite{vanderziel:1965,pershan:1966}. The light-matter coupling, $\tilde{\alpha}$, is given by the magnetic Raman tensor, which describes the change in electronic polarizability by the magnon, $\tilde{\alpha}_{ijk}=\partial \varepsilon_{jk}/\partial M_i$. $\textbf{E}=(0,E_1(t),E_2(t))$ is the circularly polarized electric field component of the laser pulse in the $yz$-plane, 
\begin{equation}
 \textbf{E}(t) = E_0 \exp\left(-\frac{t^2}{2\sigma^2} \right) \left(0,\cos(\omega_i t),\pm \sin(\omega_j t) \right).
 \label{eq:magnonE}
\end{equation}
For circular polarization, $\omega_1=\omega_2\equiv\omega_0$, $\textbf{E}(t) \times \dot{\textbf{E}}(t) = \pm \omega_0 \mathcal{E}^2(t) \hat{\textbf{x}}$, meaning the effective magnetic field is induced parallel to the wave vector of the incident pulse. Moreover, left- and right-handed polarization waves induce magnetic fields of opposite signs, as expected from the inverse Faraday effect.  

The Fourier transform of the effective magnetic field given by Eq.~\eqref{eq:B1} can be written as
\begin{equation}
    B_x(\omega) = \frac{1}{\sqrt{2\pi}}\int\limits_{-\infty}^\infty \alpha_{xjk}(\omega,\omega')E_j(\omega-\omega')E_k(\omega')d\omega',
\label{eq:B_general_frequency}
\end{equation}
where the coupling coefficient is given by 
\begin{equation}
\alpha_{xjk}(\omega,\omega') = \frac{\tilde{\alpha}}{2}\epsilon_{xjk}i(2\omega'-\omega).
\end{equation}
Plugging Eq.~\eqref{eq:B_general_frequency} into Eq.~\eqref{eq:M_general_frequency}, we can write the induced magnetization arising from the magnon dynamics as
\begin{equation}
\label{eq:magnetization}
    M_x(\omega) = \frac{1}{\sqrt{2\pi}}\int\limits_{-\infty}^\infty \chi^{(2)}_{\text{me},xjk}(\omega,\omega')E_j(\omega-\omega')E_k(\omega')d\omega',
\end{equation}
where $\chi^{(2)}_{\text{me},xjk}(\omega,\omega')$ takes the form of a second-order nonlinear magneto-electric susceptibility that contains the light-matter coupling,
\begin{align}
\chi^{(2)}_{\text{me},xjk}(\omega,\omega') =& \frac{2\hbar\gamma_{el}^2 S}{V_c\left(1+\kappa_{el}^2\right)} \frac{i\kappa_{el}\omega+2SD_y/\hbar}{\Delta_m(\omega)} \nonumber \\
&\times \alpha_{xjk}(\omega,\omega').
\label{eq:X_general_frequency}
\end{align}
For the two-photon mechanism specifically, we find that Eq.~\eqref{eq:B_general_frequency} is explicitly given by 
\begin{equation}
B(\omega) = \pm \frac{\tilde{\alpha}\sigma \omega_0 E_0^2}{\sqrt{2}} \exp\left( -\frac{\sigma^2\omega^2}{4}\right). 
\end{equation}
This amplitude is centered around $\omega=0$, which makes the induced magnetization and magnetic field quasistatic. Circularly polarized light therefore does not produce a sum-frequency component of the effective magnetic field \cite{Juraschek2019}. The same will be true for the remaining mechanisms when the two exciting particles have the same natural frequency and are circularly polarized.

Similar to the electro-phononic mechanisms, we will study the susceptibility of each mechanism through the average given by
\begin{equation}
 \chi_{\mathrm{me},xjk}^{(2)}(\omega) \equiv \frac{1}{\Delta \omega} \int_{\Delta \omega} \chi_{\mathrm{me},xjk}^{(2)}(\omega,\omega') d\omega'
\label{eq:Xme_av}
\end{equation}
over a specified frequency range $\Delta \omega$ that captures all relevant features.


\subsection{Two-phonon excitation}

Analogously to the two-phonon excitation mechanisms in the electro-phononic effects, coherently excited IR-active phonons can replace photons in the scattering process with magnons, leading to IRS \cite{nova:2017,Juraschek2020_3}, SF-IRS \cite{Juraschek2021}, and a displacive excitation \cite{kahana2023lightinduced}, as listed in Table~\ref{tab:orderofexcitations}. Instead of circularly polarized light, coherent circularly polarized, or chiral, phonons provide the angular momentum in the scattering process with the magnons. The coupling in Eq.~\eqref{eq:generalcoupling} contains the amplitudes of the driven IR-active chiral phonon mode, $A_i\equiv Q_i$, and $a$ again is an antisymmetric coupling tensor, here containing the spin-phonon coupling. The effective magnetic field produced by the phonons can therefore be written as

\begin{equation}
\label{eq:phononIFE}
 \textbf{B}(t) = \frac{\mu_0 \gamma_{ph}}{V_c}\textbf{Q}(t) \times \dot{\textbf{Q}}(t), 
\end{equation}
where $\gamma_{ph}$ is the gyromagnetic ratio of the chiral phonon, $\mu_0$ is the vacuum permeability, and $V_c$ is the unit cell volume. $\textbf{Q} \times \dot{\textbf{Q}}$ is
the phonon angular momentum, where $\textbf{Q} = (0, Q_1, Q_2)$, and $Q_1$ and $Q_2$
are the phonon amplitudes to the two orthogonal components of a doubly degenerate phonon mode in the $yz$-plane, with effective charges $\textbf{Z}_1=(0,Z_0,0)$ and $\textbf{Z}_2=(0,0,Z_0)$. A circular superposition of these components results in a chiral phonon mode. The generation of an effective magnetic field by a chiral phonon mode is known as the phonon inverse Farday, or phonon Barnett effect \cite{Juraschek2020_3,Geilhufe2023,Luo2023,Basini2024,Davies2024,Romao2024_NV,Nielson2023}. The phonon gyromagnetic ratio can obtain values spanning several orders of magnitude, from fractions of a nuclear gyromagnetic ratio to several times the electron gyromagnetic ratio, arising from various microscopic mechanisms, including ionic charge currents, electron-phonon, spin-phonon, and orbit-lattice coupling \cite{nova:2017,juraschek2:2017,Shin2018,Juraschek2019,Juraschek2020_3,Geilhufe2021,Juraschek2022_giantphonomag,Xiong2022,Geilhufe2023,Basini2024,Davies2024,Luo2023,Nielson2023,Zhang2023chiral,Gao2023,kahana2023lightinduced,Chaudhary2023,Merlin2023,Tang2024}.

The response of the chiral phonon mode is given by the  equations of motion 
\begin{align}
    \ddot{Q}_1 & + \kappa_0 \dot{Q}_1 + \Omega_0^2{Q}_1 = Z_0  E_1(t) \nonumber\\
    &+ \frac{\mu_0\gamma_{el}\gamma_{ph}}{V_c}\left( 2\dot{Q}_2(S_{1,x}+S_{2,x}) + {Q}_2(\dot{S}_{1,x}+\dot{S}_{2,x}) \right), \label{eq:phononeom1}\\
    \ddot{Q}_2 & + \kappa_0 \dot{Q}_2 + \Omega_0^2{Q}_2 = Z_0  E_2(t) \nonumber\\
    &- \frac{\mu_0\gamma_{el}\gamma_{ph}}{V_c}\left( 2\dot{Q}_1(S_{1,x}+S_{2,x}) + {Q}_1(\dot{S}_{1,x}+\dot{S}_{2,x}) \right), \label{eq:phononeom2}
\end{align}
from which we can derive analytical expressions of the phonon amplitudes to first order in the electric field component of the laser
pulse given by Eq.~\eqref{eq:magnonE},
\begin{equation}
    Q_1(\omega) = Z_0 \frac{E_1(\omega)}{\Delta_{0}(\omega)}, ~ Q_2(\omega) = Z_0 \frac{E_2(\omega)}{\Delta_{0}(\omega)}
\label{eq:QyQz_chiral_frequency}
\end{equation}
where $\Delta_{0}(\omega) = \Omega_0^2-\omega^2+i\kappa_0\omega$, and $\Omega_0\equiv \Omega_1=\Omega_2$ is the eigenfrequency of the doubly degenerate phonon mode in the $yz$-plane and $\kappa_0\equiv\kappa_1=\kappa_2$ is its linewidth. 

Using Eq. \eqref{eq:QyQz_chiral_frequency} for the phonon amplitudes, the effective magnetic field, magnetization, and magneto-electric susceptibility in Eqs.~\eqref{eq:B_general_frequency}, \eqref{eq:magnetization}, and \eqref{eq:X_general_frequency} contain the interaction of the phonons with the spins, 
\begin{equation}
    \alpha_{xjk}(\omega,\omega') = \frac{\mu_0\gamma_{ph}Z^2_0\epsilon_{xjk}i(2\omega'-\omega)}{2V_c\Delta_{0}(\omega-\omega')\Delta_{0}(\omega')}  . 
\end{equation}


\subsection{One-photon-one-phonon excitation}

Similar to the one-photon-one-phonon excitations in the electro-phononic effects, the scattering processes for magnons combine features from the two-photon and two-phonon excitation mechanisms by inducing changes in the electronic and magnonic contributions to the polarizability. Accordingly, equivalent mechanisms of IRRS and SF-IRRS, as well as a displacive mechanism should be possible, but have neither been described theoretically, nor experimentally, to our knowledge. An overview of the scattering processes can again be found in Table~\ref{tab:orderofexcitations}.
The coupling in Eq.~\eqref{eq:generalcoupling} here contains the electric field component of light $A_1\equiv E_i$ and the amplitude of a driven IR-active phonon mode, $A_2\equiv Q_2$. The effective magnetic field produced by the circular superposition of the two fields can be written as
\begin{equation}
    \textbf{B}(t) 
    = \eta \textbf{Q}(t) \times \textbf{E}(t),
\end{equation}
where $\eta$ is a coupling constant arising from photon-phonon-spin coupling. This mechanism can be seen as a hybrid magneto-opto-phononic inverse Faraday effect. Since there is no precedent for this mechanism in literature, the magnitude of $\eta$ is unknown. We can nevertheless derive the nonlinear magneto-electric susceptibility. 

The response of the chiral phonon mode is given by the following equations of motion: 
\begin{align}
    \ddot{Q}_1  + \kappa_0 \dot{Q}_1 + \Omega_0^2{Q}_1 =& Z_0  E_1(t) \nonumber \\
    &+ \eta\gamma_{el}E_2(t)(S_{1,x}+S_{2,x}), \label{eq:phononeom1_}\\
    \ddot{Q}_2  + \kappa_0 \dot{Q}_2 + \Omega_0^2{Q}_2 =& Z_0  E_2(t) \nonumber \\
    &- \eta\gamma_{el}E_1(t)(S_{1,x}+S_{2,x}). \label{eq:phononeom2_}
\end{align}
Up to first order in the electric field, the solution for the phonon amplitudes is also given by Eq. \eqref{eq:QyQz_chiral_frequency}.

The effective magnetic field, magnetization, and magneto-electric susceptibility in Eqs.~\eqref{eq:B_general_frequency}, \eqref{eq:magnetization}, and \eqref{eq:X_general_frequency} contain the photon-phonon-spin interaction, 
\begin{equation}
    \alpha_{xjk}(\omega,\omega') = \frac{\eta Z_0 \epsilon_{xjk}}{2}\left( \frac{1}{\Delta_{0}(\omega-\omega')}-\frac{1}{\Delta_{0}(\omega')} \right).
\end{equation}
The hybrid mechanism is anticipated to yield comparable effects to those observed in two-photon excitation, given that its duration will be constrained by the lifetime of the photon within the ultrashort pulse.


\begin{table}[t]
\centering
\def\arraystretch{1.3}
\caption{\textbf{Parameters for magneto-phononic effects.} $J$ is the antiferromagnetic exchange interaction,
and $D_x$ and $D_y$ are the anisotropy energies. $\gamma_{el}$ and $\gamma_{ph}$ are electron and phonon gyromagnetic ratios, respectively. $Z_0$ is the mode effective charge of the chiral phonon. $\Omega_i$ and $\kappa_i$ is the eigenfrequency and linewidth, respectively, of phonon mode $i\in\{1,2\}$. For a degenerate phonon mode, $i\equiv 0$. $V_c$ is the volume of the unit cell. $\Omega_m=\frac{2}{\hbar}\sqrt{\frac{(J+D_x)D_y}{1+\kappa_{el}^2}}$ is the damping-renormalized magnon frequency, and $\kappa_m=\frac{2\kappa_{el}}{\hbar\left(1+\kappa_{el}^2\right)}\left(J+D_x+D_y\right)$ is the magnon linewidth. 
}
\begin{tabular}{cccc}
\hline\hline
Param. & Value & Param. & Value  \\ 
\hline
$J$ & 106 meV & $D_x$ & 3.5 $\mu$eV \\
$D_y$ & 160 $\mu$eV & $\gamma_{el}$ & $-2\times 10^{11}$ $\frac{\text{C}}{\text{kg}}$ \\
$\gamma_{ph}$ & $10^{11}$ $\frac{\text{C}}{\text{kg}}$ & $Z_0$ & $1$ $\frac{e}{\sqrt{u}}$ \\ 
$V_c$ & $35$ $\text{\AA}^3$ & $\Omega_m/(2\pi)$ &  $2$ THz   \\  
$\tilde{\alpha}$ & $5\times 10^{-5}$ $\frac{\mu_0}{V_c}$ $\frac{e^3\mathrm{ps}^4}{u^2}$ & $\eta$  & 200 $\frac{\mu_0}{V_c}$ $\frac{e^2\mathrm{ps}}{u^{3/2}}$ \\
$\kappa_{el}$ & $2.4\times10^{-4}$ & $\kappa_i$ & $0.05\frac{\Omega_i}{2\pi}$ \\
\hline\hline
\end{tabular}
\label{tab:parameters2}
\end{table}


\subsection{Examples of magneto-phononic magnetizations and susceptibilities}

\begin{figure*}[t] 
    \centering


    

    
    
    \includegraphics[width=0.9\textwidth]{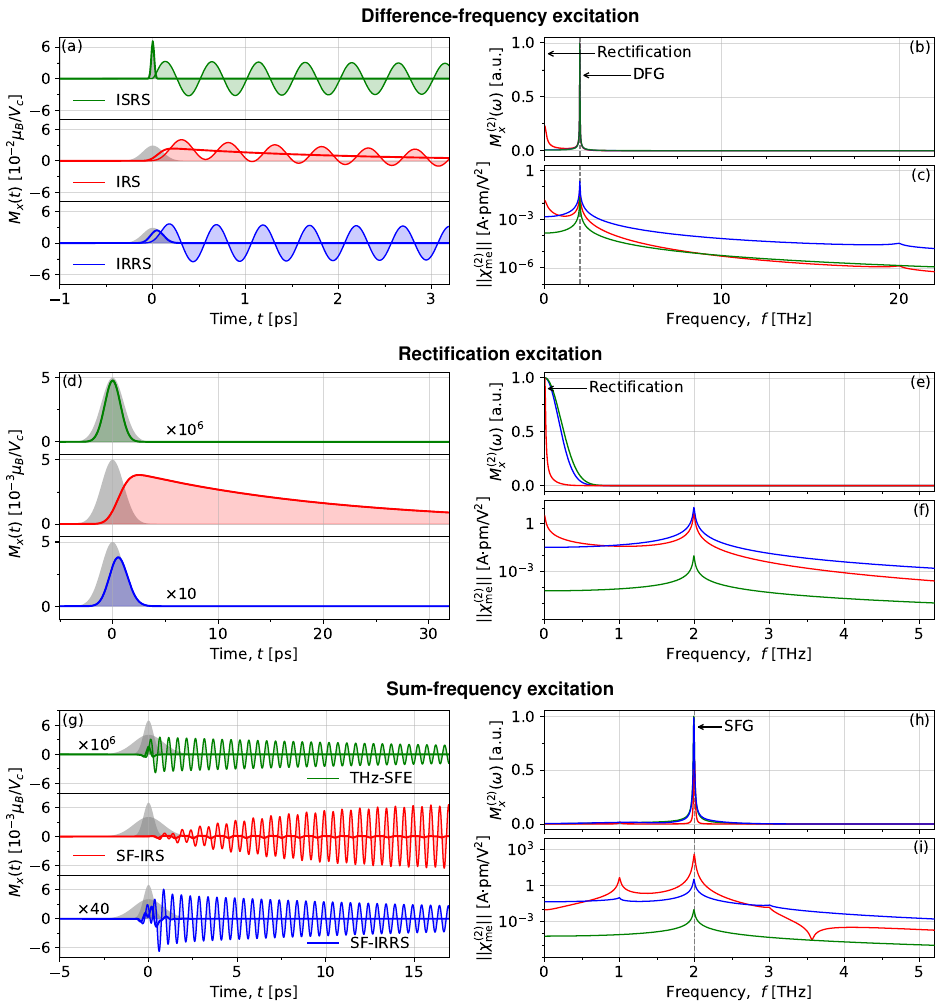}
    \caption{
    \textbf{Magneto-phononic effects at the example of a 2~THz magnon mode.} 
    (a--c) Demonstration of rectification and difference-frequency generation according to ISRS, IRS, and IRRS. 
    We choose a wavelength of 800~nm (with peak electric field $E_0=20$~MV/cm and pulse duration $\tau=50$~fs) for the two-photon process and $\omega_i/(2\pi)=\Omega_i/(2\pi)=10$~THz ($E_0=10$~MV/cm, $\tau=200$~fs) for the two-phonon and one-photon-one-phonon processes. 
    (a) Time evolution of the magnetizations for each of the three mechanisms. The pulse duration is indicated as grey shaded area.
    (b) Normalized Fourier transforms of the time traces in (a). 
    (c) Norm of the averaged nonlinear magneto-electric susceptibilities. 
    (d--f) We take the frequencies of the exciting particles to be smaller than the magnon mode frequency, $\Omega_i,\omega_i < \Omega_m$. Specifically, $\omega_i/(2\pi) = \Omega_i/(2\pi)=1$ THz. We use a circularly polarized electric field with parameters $E_0 = 0.1$ MV/cm and $\tau = 2.5$ ps. 
    %
    (d) Time evolution of the magnetizations for each of the three mechanisms, excited solely via rectification. 
    (e) Normalized Fourier transforms of the time traces in (d). 
    (f) Norm of the averaged nonlinear magneto-electric susceptibilities. 
    (g--i) Again, we take the frequencies of the exciting particles to be smaller than the magnon mode frequency, $\Omega_m > \Omega_1, \Omega_2$. However, we now consider a nondegenerate system with $\Omega_1 \neq \Omega_2$. Specifically, $\Omega_1/(2\pi) = 0.5$ THz and $\Omega_2/(2\pi) = 1.5$ THz. We use an electric field with parameters $E_0 = 0.05$ MV/cm, $\tau = 1$ ps, and $\omega_0/(2\pi) = 1$ THz as a reference. We adjust the peak electric field and pulse duration so that the energy remains constant and excite the two phonons with frequencies $\omega_1/(2\pi) = 0.5$ THz and $\omega_2/(2\pi) = 1.5$ THz, respectively.
    (g) Time evolution of the magnetizations for each of the three mechanisms, excited by nondegenerate phonons via SFG. 
    (h) Normalized Fourier transforms of the time traces in (g). 
    (i) Norm of the averaged nonlinear magneto-electric susceptibilities.
    }
    \label{fig:main2}
\end{figure*}

As for the electro-phononic effects, we now evaluate the formalism developed in the previous sections for the three magneto-phononic mechanisms, using typical values for the magnon, phonon, and light-matter interactions parameters, which we display in Table~\ref{tab:parameters2}. For the one-phonon-one-photon hybrid effect, for which there are no precedents in literature, we choose the coupling coefficient such that they lead to results of similar magnitude to the others. In contrast to the dynamics observed in a system featuring a coupled phonon, the excitation of a magnon through the scattering of two particles with identical natural frequencies lacks a sum-frequency component. 
In Fig.~\ref{fig:main2}, we show the time evolutions of the magnetizations arising from the out-of-plane alignment of the spins, $M_x(t)$, according to the three mechanisms. We distinguish two cases: In Fig.~\ref{fig:main2}(a--c), we take the frequencies of the exciting particles to be larger than the magnon frequency, $\omega_i, \Omega_i > \Omega_m$, with a wavelength of 800~nm for the two-photon process and $\omega_i/(2\pi)=\Omega_i/(2\pi)=10$~THz for the two-phonon and one-photon-one-phonon processes, in order to illustrate magneto-phononic difference-frequency generation and rectification. In Fig.~\ref{fig:main2}(d--f), we reduce the frequency of the exciting particles to $\omega_i/(2\pi)=\Omega_i/(2\pi)=1$~THz, while keeping the number of cycles $\omega_0\tau$ of the pulse
and pulse energy constant, to illustrate the absence of sum-frequency generation. 
The two-photon mechanisms are plotted in green, the two-phonon mechanisms in red, and the one-photon-one-phonon mechanisms in blue. We use the analytical solutions of $M_x$ given by Eq.~\eqref{eq:Mx_time_domain} in the Appendix. The analytical solutions match remarkably well with the numerical evaluations of the respective equations of motion, as shown in Fig.~\ref{fig:analyticalcomparisonmagnons} in the Appendix.



Fig.~\ref{fig:main2}(a) displays the temporal evolution of the magnetizations for the three difference-frequency processes. The excitation induced by the difference-frequency components of each of the mechanisms causes an impulsive response of the magnetization component of the magnon. As for the electro-phononic effects, our analysis assumes that the excitation energy lies well below the band-gap energy. Consequently, for mechanisms directly involving the electric field component of light, ISRS and IRRS, rectification occurs solely during the duration of the laser pulse, which here is less than one oscillation period of the magnon. IRS, as in the electro-phononic mechanism, in turn induces a prolonged rectified state as long as the driven phonon modes continue ringing. In Fig.~\ref{fig:main2}(b), we depict the normalized Fourier transforms of the temporal traces illustrated in (a). Distinct and symmetrical DFG peaks emerge at the eigenfrequency of the coupled magnon at 2~THz. Additionally, a static component is observed at zero frequency for IRS, indicating the unidirectional force produced by the effective magnetic field acting on the coupled magnon. For ISRS and IRRS, the static components are nonzero, but negligible. In Fig.~\ref{fig:main2}(c), we plot the norm of the average of the nonlinear magneto-electric susceptibility given by Eq.~\eqref{eq:Xme_av}. Here, the features of rectification and DFG can be identified for IRS, whereas ISRS and IRRS show primarily DFG peaks at the eigenfrequency of the magnon.

Fig.~\ref{fig:main2}(d) shows the temporal evolution of the magnetization component of the magnon for the three processes. Notably, the magnetization shows only rectification, but no oscillatory response, and therefore no sum-frequency or second-harmonic generation. In contrast to the electro-phononic effects, the rectification component is dominant here. Accordingly, the normalized Fourier transforms of the time traces plotted in Fig.~\ref{fig:main2}(e) only show components centered around zero frequency, corresponding to a quasistatic magnetization. In Fig.~\ref{fig:main2}(f), we finally show the norm of the average of the nonlinear magneto-electric susceptibility, which shows a clear peak at zero frequency for IRS, corresponding to rectification. All three susceptibilities again show peaks at the eigenfrequency of the magnon.

While for degenerate phonon modes, there is no SFG/SHG component, for nondegenerate phonons, with $\Omega_1 \neq \Omega_2$ and $\Omega_{1/2}<\Omega_m$, there is. In Fig.~\ref{fig:main2}(g), we therefore show the temporal evolutions of the magnetizations, which now arises as a result of excitation via the SFG component of each mechanism. 
Consequently, we obtain a gradual increase in magnetization, instead of the usual abrupt increase observed with excitation via DFG. The magnetization spectrum in Fig. \ref{fig:main2}(h) now shows peaks corresponding to SFG, instead of rectification and DFG. These features are also evident in the magnetoelectric susceptibility displayed in Fig. \ref{fig:main2}(i).

All calculations performed in Fig.~\ref{fig:main2} are continuously scalable to higher pulse fluences. As discussed in the electro-phononic effects section, the amount of induced phonon amplitude and therefore magnetization of the magnon is limited by the stability of the crystal lattice, described by the Lindemann criterion \cite{Lindemann1910,sokolowski-tinten:2003}. The maximum amplitudes of the IR-active phonon modes, shown in Fig.~\ref{fig:Qd2} in the Appendix, are on the order of $Q\approx 2$~\AA$\sqrt{u}$, which is typically below this limit \cite{Juraschek2018}. The relative strengths of the two-photon, two-phonon, and one-photon-one-phonon processes are strongly material dependent and are evaluated here for typical parameters found in literature, where applicable (Table~\ref{tab:parameters2}).


\subsection{2D nonlinear magneto-electric susceptibility}

\begin{figure}[t]
    \centering
      
      
    \includegraphics[width=1.\linewidth]{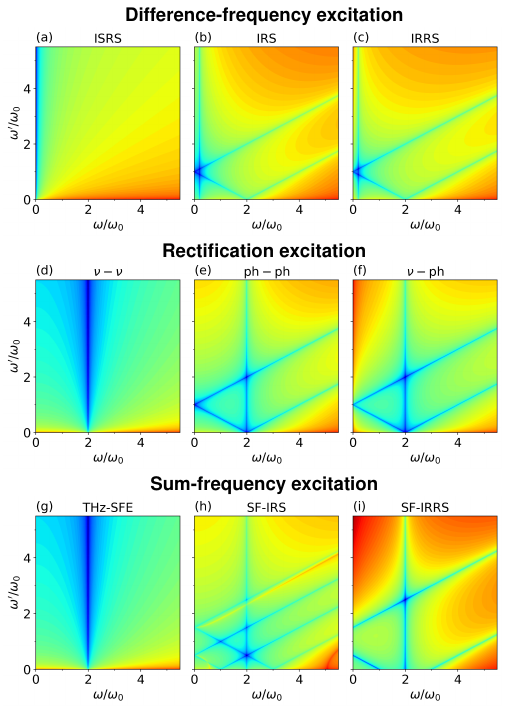}
    \caption{
    \textbf{2D nonlinear magneto-electric susceptibility.} We show the norm of the second-order nonlinear magneto-electric susceptibility, $||\chi_{\text{me},xyz}(\omega,\omega')||$ on a logarithmic scale for the three difference-frequency (a--c), rectification (d--f), and sum-frequency 
    (g--i) excitation mechanisms. More intense blue indicates larger values, more intense red indicates smaller values, plotted in arbitrary units.
    }
    \label{fig:HM2}
\end{figure}

In Fig.~\ref{fig:HM2}, we display the norm of the magneto-electric susceptibility denoted as $||\chi_{\text{me},xyz}(\omega,\omega')||$. Within Fig.~\ref{fig:HM2}(a--c), the frequencies of the stimulating particles exceed the magnon frequency, $\omega_i, \Omega_i > \Omega_m$. This scenario leads to excitation via difference-frequency generation and rectification. The magnon mode has an eigenfrequency of $\Omega_m/(2\pi) = 2$~THz, while we set the wavelength to 800~nm for the two-photon process and $\Omega_i/(2\pi)=10$~THz for the two-phonon and one-photon-one-phonon processes. (a), (b), and (c) correspond to ISRS, IRS, and IRRS, respectively. In Fig.~\ref{fig:HM2}(d--f), the frequencies associated with the exciting particles are half of that of the magnon frequency, $\Omega_i = \Omega_c/2$. With a magnon frequency of $\Omega_m/(2\pi) = 2$~THz, we set $\omega_i/(2\pi)=\Omega_i/(2\pi)=1$~THz for all three mechanisms. Due to the nature of the mechanism, THz-SFE (a), SF-IRS (b), and SF-IRRS (c) only lead to rectification, but not sum-frequency or second-harmonic generation. 
Finally, in Fig. \ref{fig:HM2}(g--i), the frequencies associated with the stimulating particles are again smaller than the magnon frequency, $\Omega_{1/2} < \Omega_c$. However, now we introduce nondegenerate phonons with $\Omega_1 \neq \Omega_2$, resulting in excitation via sum-frequency generation. Here, $\Omega_1/(2\pi)=0.5$ THz, and $\Omega_2/(2\pi)=1.5$ THz.


\section{Conclusion}\label{sec:discussion}



Our theoretical framework enables a unified description of three-particle, Raman-type scattering processes by Raman-active phonons and magnons, for which the three-particle interaction vortex can be mediated by infrared-active phonons. The electro-phononic processes so-enabled in noncentrossymmetric materials can be regarded as phonon-mediated electro-optic effects, whereas the magneto-phononic processes in antiferromagnetic materials can be seen as phonon-mediated magneto-optic effects. The framework captures a broad variety of existing scattering processes, summarized in Table~\ref{tab:orderofexcitations}, and in addition predicts a hybrid magneto-opto-phononic inverse Faraday effect that can be measured in state-of-the-art pump-probe setups, for example using Faraday rotation or magneto-optic Kerr effect (MOKE) measurements. A particular feature of the two-phonon excitation mechanisms is the generation of phononic and magnonic rectification even for photon energies far below the band gap of the material, avoiding the strong dissipation inherent to excited electronic states. These enable the creation of nonequilibrium crystal and spin structures that are not accessible in equilibrium.

Our discussion has centered on mechanisms that involve coherently driven IR-active phonons as transducers for the scattering processes with the Raman-active phonons or magnons. In recent years, an increasing number of scattering processes has further been demonstrated that involves coherently driven magnons as transducers \cite{Lu2017,Mashkovich2021,Juraschek2021_5,Blank2023_2,Zhang2024_mixing,Schoenfeld2023,Zhang2024_upconversion,Huang2024}. Our formalism is readily extendable to these processes, leading equivalently to electro-magnonic and magneto-magnonic processes, which will be done in a future study. 


\section*{Acknowledgments}

We acknowledge T. Kahana, M. Udina, and F. Barantani for useful discussions. D.A.B.L. and W.H. acknowledge support from the U.S. Department of Energy, Office of Science, Office of Basic Energy Sciences Early Career Research Program under Award Number DE-SC-0021305. D.M.J. acknowledges support from Tel Aviv University.



\section*{Appendix A: Special Functions}

Before deriving the analytical solutions of the equations of motion presented in the main text, we define a variety of functions and integrals
Let's consider the following functions in frequency domain
\begin{subequations}
\begin{align}
F(\omega) &= \frac{1}{\Omega^2-\omega^2-i\kappa\omega}, \\
G(\omega) &= \frac{i\omega}{\Omega^2-\omega^2-i\kappa\omega},
\end{align}
\end{subequations}
with their corresponding representation in time domain
\begin{subequations}
\begin{align}
F(t)= & \sqrt{2\pi}\frac{e^{-\kappa t/2}}{\widetilde{\Omega}}\sin\left(\widetilde{\Omega}t \right)\theta(t), \\
G(t)= & \sqrt{2\pi}e^{-\kappa t/2}\left[\cos\left(\widetilde{\Omega}t\right) -\frac{\kappa}{2\widetilde{\Omega}}\sin\left(\widetilde{\Omega}t\right)\right]\theta(t), 
\end{align}
\end{subequations}
where $\widetilde{\Omega}=\sqrt{\Omega^2-\frac{\kappa^2}{4}}$ and $\theta(t)$ is the Heaviside step function. Let's define
\begin{subequations}
\begin{align}
f_1(t) &= 1+\text{erf}\left(  \frac{2t+\sigma^2\left( 2i(\omega_0-\widetilde{\Omega})-\kappa\right)}{2\sqrt{2}\sigma}  \right), \\
f_2(t) &= 1+\text{erf}\left(  \frac{2t+\sigma^2\left( 2i(\omega_0+\widetilde{\Omega})-\kappa\right)}{2\sqrt{2}\sigma}  \right), \\
\Theta_1(t) &= \widetilde{\Omega}t + \frac{\sigma^2 \kappa}{2}\left(\omega_0-\widetilde{\Omega} \right), \\
\Theta_2(t) &= -\widetilde{\Omega}t + \frac{\sigma^2 \kappa}{2}\left(\omega_0+\widetilde{\Omega} \right), \\
A(t) &= \exp\left( \frac{\sigma^2}{2}\left(\frac{\kappa^2}{4}-\left(\widetilde{\Omega}+\omega_0 \right)^2 \right)-\frac{\kappa t}{2}\right), \\
C &= \exp\left( 2\sigma^2\widetilde{\Omega}\omega_0 \right).
\end{align}
\end{subequations}

We use the previous functions to express the following integrals:
\begin{subequations}
\begin{align}
\begin{split}
I^{(1)}(t) &= \int\limits_{0}^{\infty}E_c(t-t')e^{-\kappa t'/2}\cos\left(\widetilde{\Omega}t'\right)dt' \\
&= \frac{E_0\sigma\sqrt{\pi}A(t)}{2\sqrt{2}} \text{Re}\left[ Cf_1(t)e^{-i\Theta_1(t)}+f_2(t)e^{-i\Theta_2(t)}  \right], 
\end{split}\label{eq:specialI1} \\
\begin{split}
I^{(2)}(t) &= \int\limits_{0}^{\infty}E_c(t-t')e^{-\kappa t'/2}\sin\left(\widetilde{\Omega}t'\right)dt' \\
&= \frac{E_0\sigma\sqrt{\pi}A(t)}{2\sqrt{2}}\text{Im}\left[ -Cf_1(t)e^{-i\Theta_1(t)}+f_2(t)e^{-i\Theta_2(t)}  \right], 
\end{split}\label{eq:specialI2} \\
\begin{split}
I^{(3)}(t) &= \int\limits_{0}^{\infty}E_s(t-t')e^{-\kappa t'/2}\cos\left(\widetilde{\Omega}t'\right)dt' \\
&= \frac{E_0\sigma\sqrt{\pi}A(t)}{2\sqrt{2}}\text{Im}\left[ -Cf_1(t)e^{-i\Theta_1(t)}-f_2(t)e^{-i\Theta_2(t)}  \right], 
\end{split}\label{eq:specialI3} \\
\begin{split}
I^{(4)}(t) &= \int\limits_{0}^{\infty}E_s(t-t')e^{-\kappa t'/2}\sin\left(\widetilde{\Omega}t'\right)dt' \\
&= \frac{E_0\sigma\sqrt{\pi}A(t)}{2\sqrt{2}}\text{Re}\left[ Cf_1(t)e^{-i\Theta_1(t)}-f_2(t)e^{-i\Theta_2(t)}  \right]. 
\end{split}\label{eq:specialI4}
\end{align}
\label{eq:integralsI}
\end{subequations}

\noindent{}The electric field components are given by

\begin{subequations}
\begin{align}
E_c(t) &= \mathcal{E}(t)\cos(\omega_0t), \\
E_s(t) &= \mathcal{E}(t)\sin(\omega_0t),
\end{align}
\end{subequations}
where $\mathcal{E}(t)=E_0\exp\left(-t^2/(2\sigma^2)\right)$ is the Gaussian carrier envelope of the pulse, $E_0$ is the peak electric field, $\sigma=\tau/\sqrt{8\ln2}$, and $\tau$ is the full width at half maximum pulse duration. 

From now on, $I_{\alpha,\beta}^{(\sigma)}$, $\sigma\in\{1,2,3,4\}$, will refer to any of the functions in Eqs.~\eqref{eq:integralsI}, after replacing the parameters $\Omega$ and $\kappa$ with those of the respective phonon $\alpha$, and $E_0$ and $\tau$ with those of the electric field component $E_\beta$.


\section*{Appendix B: Detailed derivations of the electro-phononic effects}

The total potential energy including all three three-particle scattering processes (two-photon, two-phonon, and one-photon-one-phonon excitations of the coupled phonon) is given by
\begin{align}
V = &~ \underbrace{ \frac{\Omega_c^2}{2}Q_c^2+\frac{\Omega_1^2}{2}Q_1^2+\frac{\Omega_2^2}{2}Q_2^2 }_{\text{Harmonic phonons}} \nonumber \\
& \underbrace{-(Q_cZ_{c,i}+Q_1Z_{1,i}+Q_2Z_{2,i})E_i}_{\text{IR absorption}} \nonumber \\
& \underbrace{-R_{ij}E_iE_jQ_c}_{\text{two-photon}} \underbrace{-cQ_cQ_1Q_2}_{\text{two-phonon}} \nonumber  \\
& \underbrace{-b_iE_iQ_2Q_c,}_{\text{one-photon-one-phonon}}
\end{align}
where $Q_\sigma$ is the normal mode coordinate (or amplitude) of phonon mode $\sigma$, $\Omega_\sigma$ is the corresponding phonon frequency, and $\textbf{Z}_\sigma$ is the mode effective charge. $\textbf{R}$ is the Raman tensor for the two-photon excitation, $c$ is the nonlinear phonon coupling coefficient for the two-phonon excitation, and $\textbf{b}$ is the coupling coefficient for the one-photon-one-phonon excitation. $\textbf{E}$ is the electric field component of the laser pulse. Because $\textbf{Z}_c$ is nonzero in noncentrosymmetric materials, the Raman-active phonon mode could in principle couple to the laser pulse directly. Generally, we look at cases in which the center frequency of the laser pulse is far off resonance from the eigenfrequency of this phonon mode, and we therefore neglect this coupling term in all calculations. Consequently, the primary excitation of the coupled Raman-active phonon comes from the three-particle scattering processes.

The equations of motion connected to the above phonon potential can be written as
\begin{subequations}
\begin{align}
\begin{split}
\ddot{Q}_c + \kappa_c\dot{Q}_c+\Omega_c^2Q_c = &~ Z_{c,i}E_i(t) + R_{ij}E_i(t)E_j(t) \\
& + cQ_1Q_2 + b_iQ_2E_i(t) , \label{eq:phononeom_APPENDIXc}
\end{split} \\
\ddot{Q}_1+\kappa_1\dot{Q}_1+\Omega_1^2Q_1 = &~ Z_{1,i}E_i(t)+cQ_2Q_c \label{eq:phononeom_APPENDIX1}\\
\begin{split}
\ddot{Q}_2+\kappa_2\dot{Q}_2+\Omega_2^2Q_2 = &~ Z_{2,i}E_i(t)+cQ_1Q_c \label{eq:phononeom_APPENDIX2}\\
&+b_iQ_cE_i(t),
\end{split}
\end{align}
\end{subequations}
where $\kappa_\sigma$ is the phonon linewidth of phonon mode $\sigma$.

In the frequency domain, the  solutions for the phonon amplitudes within the aforementioned equations of motion satisfy
\begin{subequations}
\begin{align}
Q_c(\omega) = &  \underbrace{\frac{Z_{c,i}}{\Delta_c(\omega)}E_i(\omega)}_{\text{IR absorption}} \nonumber\\
& + \underbrace{\frac{R_{ij}}{\Delta_c(\omega)}E_i(\omega)\circledast E_j(\omega)}_{\text{two-photon}} \nonumber \\
&+ \underbrace{\frac{c}{\Delta_c(\omega)}Q_1(\omega)\circledast Q_2(\omega)}_{\text{two-phonon}} \nonumber\\
& +\underbrace{\frac{b_i}{\Delta_c(\omega)}E_i(\omega)\circledast Q_2(\omega)}_{\text{one-photon-one-phonon}}, 
\label{eq:Qc_all} \\
Q_1(\omega) = & \underbrace{\frac{Z_{1,i}}{\Delta_1(\omega)}E_i(\omega)}_{\text{IR absorption}} \nonumber\\
& + \underbrace{\frac{c}{\Delta_1(\omega)}Q_c(\omega)\circledast Q_2(\omega)}_{\text{two-phonon backaction}}, \\
Q_2(\omega) = &  \underbrace{\frac{Z_{2,i}}{\Delta_2(\omega)}E_i(\omega)}_{\text{IR absorption}} \nonumber\\
&+\underbrace{\frac{b_i}{\Delta_2(\omega)}E_i(\omega) \circledast Q_c(\omega)}_{\text{two-photon backaction}} \nonumber \\
&+ \underbrace{\frac{c}{\Delta_2(\omega)}Q_c(\omega)\circledast Q_1(\omega)}_{\text{two-phonon backaction}}, 
\end{align}
\end{subequations}
where $\Delta_\sigma=\Omega_\sigma^2-\omega^2+i\omega\kappa_\sigma$ and $\circledast$ is the convolution operator. It is noteworthy that the terms associated with infrared absorption are of first order in the electric field, whereas the remaining terms manifest as second order or higher. The system of equations of motion can then be straightforwardly solved  perturbatively.

We begin the analysis by deriving the first-order solutions, wherein the contributions arising from the three-particle scattering processes are disregarded.

\noindent\underline{First-order solutions}
\begin{subequations}
\begin{align}
Q_c^{(1)}(\omega) &= \frac{Z_{c,i}}{\Delta_c(\omega)}E_i(\omega), 
\label{eq:Qc_1}
\\
Q_1^{(1)}(\omega) &= \frac{Z_{1,i}}{\Delta_1(\omega)}E_i(\omega), 
\label{eq:Q1_1}
\\
Q_2^{(1)}(\omega) &= \frac{Z_{2,i}}{\Delta_2(\omega)}E_i(\omega). 
\label{eq:Q2_1}
\end{align}
\end{subequations}
As previously mentioned, the selection of the light pulse is designed to drive the IR-active phonons $1$ and $2$ while intentionally avoiding spectral overlap with the coupled Raman-active phonon $c$. Consequently, $Q_1^{(1)}$ and $Q_2^{(1)}$ are the dominant terms for phonon modes 1 and 2. In contrast, $Q_c^{(1)}$ is negligibly small. Hence, we determine the second-order term in the solution for $Q_c$ perturbatively. Substituting Eqs.~\eqref{eq:Q1_1} and \eqref{eq:Q2_1} into Eq. \eqref{eq:Qc_all}, we find the second-order solutions.

\noindent\underline{Second-order solutions}
\begin{align}
Q_c^{(2)}(\omega) = &~ \frac{R_{ij}}{\Delta_c(\omega)}\left(E_i(\omega)\circledast E_j(\omega)\right) \nonumber \\
& +\frac{c}{\Delta_c(\omega)}\left(Q_1^{(1)}(\omega)\circledast Q_2^{(1)}(\omega)\right) \nonumber \\
&+\frac{b_i}{\Delta_c(\omega)}\left(E_i(\omega) \circledast Q_2^{(1)}(\omega)\right) \nonumber \\
&= \frac{R_{ij}}{\Delta_c(\omega)}\left(E_i(\omega)\circledast E_j(\omega)\right) \nonumber \\
& + \frac{cZ_{1,i}Z_{2,j}}{\Delta_c(\omega)}\left( \frac{E_i(\omega)}{\Delta_1(\omega)} \circledast \frac{E_j(\omega)}{\Delta_2(\omega)}\right) \nonumber \\
& + \frac{b_i Z_{2,j}}{\Delta_c(\omega)}\left(E_i(\omega)\circledast \frac{E_j(\omega)}{\Delta_2(\omega)}\right).
\end{align}
Therefore, the three-particle scattering processes induce a second-order polarization arising from the Raman-active phonon that is given by

\begin{equation}
\begin{aligned}
P_{c,i}^{(2)}(\omega) =& \frac{Z_{c,i}}{V_c}Q_c^{(2)}(\omega) \\
=& \frac{1}{\sqrt{2\pi}V_c\Delta_c(\omega)} \\
& \times \int\limits_{-\infty}^{\infty}\bigg[ \underbrace{Z_{c,i}R_{jk}}_{\text{two-photon}} +\underbrace{\frac{cZ_{c,i}Z_{1,j}Z_{1,k}}{\Delta_1(\omega-\omega')\Delta_2(\omega')}}_{\text{two-phonon}}  \\
&+\underbrace{\frac{Z_{c,i}b_jZ_{2,k}}{\Delta_2(\omega')}}_{\text{one-photon-one-phonon}}\bigg]E_j(\omega-\omega')E_k(\omega')d\omega' \\
=& \frac{\varepsilon_0}{\sqrt{2\pi}}\int\limits_{-\infty}^{\infty}\chi_{e,ijk}^{(2)}(\omega,\omega')E_j(\omega-\omega')E_k(\omega')d\omega',
\end{aligned}
\end{equation}
where $V_c$ is the unit-cell volume of the crystal, $\varepsilon_0$ is the vacuum permittivity, and we have defined the electric susceptibility as
\begin{equation}
\begin{split}
\varepsilon_0 & V_c  \chi_{e,ijk}^{(2)}(\omega.\omega') = \underbrace{\frac{Z_{c,i}\left(R_{jk}+R_{kj}\right)}{2\Delta_c(\omega)}}_{\text{two-photon}} \\
&\underbrace{+\frac{cZ_{c,i}}{2\Delta_c(\omega)}\bigg[ \frac{Z_{1,j}Z_{2,k}}{\Delta_1(\omega-\omega')\Delta_2(\omega')} + \frac{Z_{1,k}Z_{2,j}}{\Delta_1(\omega')\Delta_2(\omega-\omega')} \bigg]}_{\text{two-phonon}} \\
&+ \underbrace{\frac{Z_{c,i}}{2\Delta_c(\omega)}\left[ \frac{b_jZ_{2,k}}{\Delta_2(\omega')}+\frac{b_k Z_{2,j}}{\Delta_2(\omega-\omega')}\right]}_{\text{one-photon-one-phonon}}. \\
\end{split}
\end{equation}

Now, let us assume that the electric field is aligned along the $z$-axis, so that $\textbf{E}(t)=(0,0,E(t))$, where \begin{equation}
\begin{split}
E(t) &= \mathcal{E}(t)\cos(\omega_0t+\phi) \\
&= \mathcal{E}(t)\left[\cos(\omega_0t)\cos(\phi)-\sin(\omega_0t)\sin(\phi)\right].
\end{split}
\label{eq:GEN_E}
\end{equation}
Here, $\phi$ is the carrier envelope phase. The inverse Fourier transform of $1/\Delta_\sigma(\omega)$ is 
\begin{equation}
 \mathcal{T}_\sigma(t) = \sqrt{2\pi}\frac{e^{-\kappa_\sigma t/2}}{\widetilde{\Omega}_\sigma}\sin\left(\widetilde{\Omega}_\sigma t\right)\theta(t),
\end{equation}
where $\widetilde{\Omega}_\sigma =\sqrt{\Omega_\sigma^2-\kappa_\sigma^2/4}$ and $\theta$ is the Heaviside step function. Back to the time domain, we obtain
\begin{subequations}
\begin{align}
Q_1^{(1)}(t) =& Z_{1,z}\left(I^{(2)}_{1,1}(t)\cos(\phi_1)-I^{(4)}_{1,1}(t)\sin(\phi_1)\right. \nonumber \\
&\left.+I^{(2)}_{1,2}(t)\cos(\phi_2)-I^{(4)}_{1,2}(t)\sin(\phi_2)\right)
\nonumber \\
\approx & Z_{1,z}\left(I^{(2)}_{1,1}(t)\cos(\phi_1)-I^{(4)}_{1,1}(t)\sin(\phi_1)\right),
 \\
Q_2^{(1)}(t) =& Z_{2,z}\left(I^{(2)}_{2,1}(t)\cos(\phi_1)-I^{(4)}_{2,1}(t)\sin(\phi_1)\right. \nonumber \\
&\left.+I^{(2)}_{2,2}(t)\cos(\phi_2)-I^{(4)}_{2,2}(t)\sin(\phi_2)\right) \nonumber
\\
\approx & Z_{2,z}\left(I^{(2)}_{2,2}(t)\cos(\phi_2)-I^{(4)}_{2,2}(t)\sin(\phi_2)\right),
\end{align}
\end{subequations}
where we have assumed that we have an electric field aligned along the $z$-axis, which has contributions from two beams, $E_1$ (resonant with phonon mode 1) and $E_2$ (resonant with phonon 2), each of which has a form given by Eq.~\eqref{eq:GEN_E}. Moreover,
\begin{align}
Q_c^{(2)}(t) =& \mathcal{T}_c(t) \circledast \bigg[ \underbrace{R_{zz}E^2(t)}_{\text{two-photon}} \nonumber \\
& + \underbrace{cQ_1^{(1)}(t)Q_2^{(1)}(t)}_{\text{two-phonon}}  \nonumber \\
&+ \underbrace{b_z Q_2^{(1)}(t)E(t)}_{\text{one-photon-one-phonon}} \bigg].
\label{eq:time_Qc2}
\end{align}

For the two-phonon and one-photon-one-phonon contributions, the above result requires numerical evaluation of the convolution. However, we can obtain the explicit solution for the excitation via the two-photon mechanism. In that case, suppose we have an electric field given by Eq. \eqref{eq:GEN_E} with $\phi=0$. In the frequency domain, the phonon amplitude is given explicitly by 
\begin{align}
Q_c^{(2)}&(\omega) = \frac{R_{zz}E_0^2\sigma}{4\sqrt{2}\Delta_c(\omega)}\left[\exp\left(\frac{-\sigma^2(\omega-\omega_0)^2}{4}\right) \right. \nonumber\\
&\left. + 2\exp\left(\frac{-\sigma^2\omega^2}{4}\right) +\exp\left(\frac{-\sigma^2(\omega+\omega_0)^2}{4}\right) \right],
\label{eq:2photonamplitudefreq}
\end{align}
with a DC component equal to
\begin{equation}
Q_c^{(2)}(\omega=0) = \frac{R_{zz}E_0^2\sigma}{2\sqrt{2}\Omega_c^2}\left(1+e^{-\sigma^2\omega_0^2} \right) = \frac{R_{zz}}{\sqrt{2\pi}\Omega_c^2} \mathcal{I},
\end{equation}
where $\mathcal{I}=\int E^2(t)\mathrm{d}t$ is proportional to the intensity of the laser pulse. We can Fourier transform Eq.~\eqref{eq:2photonamplitudefreq} back to the time domain to obtain an analytical expression of $Q_c^{(2)}(t)$. If we define
\begin{equation}
\begin{split}
S_\nu(t;a) &= E_0^2\int\limits_0^\infty \exp\left(-\frac{(t-t')^2}{\sigma^2}-\frac{\kappa_\nu t'}{2}+iat'\right)\frac{dt'}{\widetilde{\Omega}_\nu} \\
&= \frac{E_0^2\sqrt{\pi}\sigma}{2\widetilde{\Omega}_\nu}\left(1+\text{erf}\left(\frac{4t+(2ia-\kappa_\nu)\sigma^2}{4\sigma}\right) \right) \\
& \times \exp\left( \frac{\sigma^2}{16}\left(\kappa_\nu^2-4a^2-4ia\kappa_\nu\right)+t\left(ia-\frac{\kappa_\nu}{2}\right)\right),
\end{split}
\label{eq:Qc_2_2photon_analytic}
\end{equation}
then
\begin{equation}
\begin{split}
Q_c^{(2)}& (t) = \\
& \frac{R_{zz}}{4}\text{Im}\bigg[\left(S_c(t;2\omega_0+\widetilde{\Omega}_c)-S_c(t;2\omega_0-\widetilde{\Omega}_c)\right)e^{-2it\omega_0}\\
&+2S_c(t;\widetilde{\Omega}_c) \bigg]. 
\end{split}
\label{eq:solQc_2_2photon_analytic}
\end{equation}


\begin{figure}[t]
\centering
\includegraphics[scale=0.61]{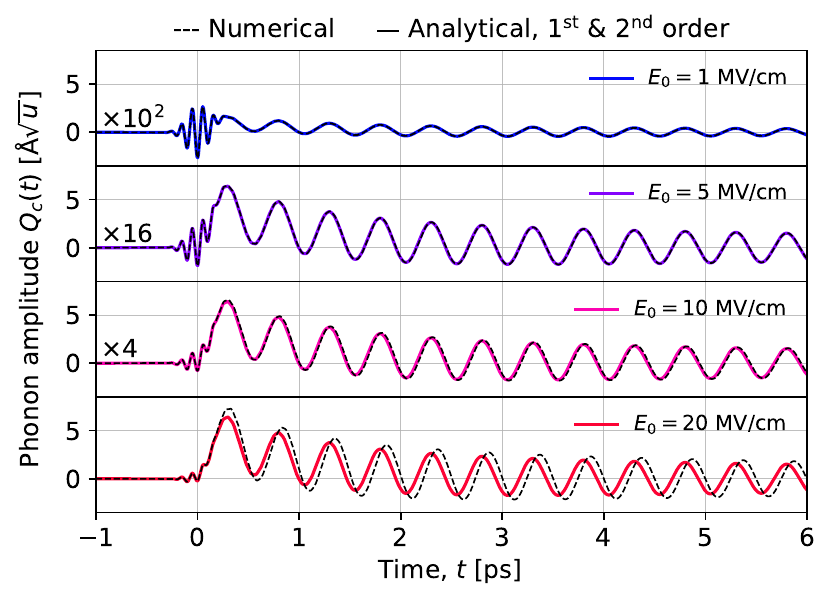}
\caption{\textbf{Comparison of analytical and numerical results for the electro-phononic effects.} Phonon dynamics as evaluated by the semi-analytical expression in Eq.~\eqref{eq:time_Qc2} compared to the numerical evaluations of the equations of motion Eqs.~\eqref{eq:phononeom_APPENDIXc}, \eqref{eq:phononeom_APPENDIX1}, and \eqref{eq:phononeom_APPENDIX2}. We show the example of ionic Raman scattering (IRS), using different amplitudes of the peak electric field. By increasing the laser intensity, it becomes evident that higher-order corrections are necessary. In this work, we only discuss contributions up to second order, and we use parameters for which the numerical results agree well with the analytical ones. 
}
\label{fig:analyticalcomparisonphonons}
\end{figure}

\begin{figure}[h]
    \centering
    
    
      
      
    
    \includegraphics[width=0.47\textwidth]{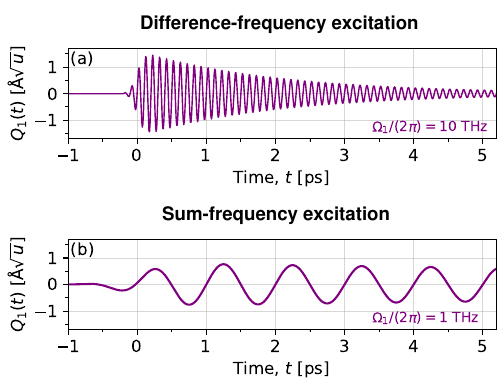}
    \caption{
    Amplitude $Q_1$ of the infrared-active phonon mode resonantly excited by an electric field, participating in the driving force of the two-phonon and one-photon-one-phonon electro-phononic mechanisms. (a) Phonon amplitude for the IRS and IRRS mechanisms in Fig.~\ref{fig:main1}(a). (b) Phonon amplitude for the SF-IRS and SF-IRRS mechanisms in Fig.~\ref{fig:main1}(d).
    }
    \label{fig:Qd1}
\end{figure}

\section*{Appendix C: Detailed derivations of the magneto-phononic effects}

\subsection*{C.1 Magnon eigenfrequencies}

We begin by deriving the eigenfrequencies of the magnon modes of the system analytically. Let $\textbf{m}=\textbf{S}_1+\textbf{S}_2$ and $\textbf{l}=\textbf{S}_1-\textbf{S}_2$, then the spin Hamiltonian can be written as
\begin{equation}
\begin{split}
   \mathcal{H} &= J \textbf{S}_1 \cdot \textbf{S}_2 + D_x\left(S_{1,x}^2+S_{2,x}^2\right)+D_y\left(S_{1,y}^2+S_{2,y}^2\right) \\
   &= \frac{J}{2}m^2 - JS^2+\frac{D_x}{2}\left( m_x^2+l_x^2\right)+\frac{D_y}{2}\left(m_y^2+l_y^2\right).
\end{split}
\end{equation}
The effective magnetic fields can thus be written as
\begin{subequations}
\begin{align}
\frac{\partial \mathcal{H}}{\partial \textbf{m}} &= (J+D_x)m_x\hat{\textbf{x}} + (J+D_y)m_y\hat{\textbf{y}} + Jm_z\hat{\textbf{z}}, \\
\frac{\partial \mathcal{H}}{\partial \textbf{l}} &= D_xl_x \hat{\textbf{x}}+ D_yl_y\hat{\textbf{y}}.
\end{align}
\end{subequations}
The equations of motion without damping are given by the Landau-Lifshitz equations
\begin{subequations}
\begin{align}
\hbar\frac{d\textbf{m}}{dt} &= -\textbf{m}\times\frac{\partial \mathcal{H}}{\partial \textbf{m}}-\textbf{l}\times\frac{\partial \mathcal{H}}{\partial \textbf{l}} ,\\
\hbar\frac{d\textbf{l}}{dt} &= -\textbf{m}\times\frac{\partial \mathcal{H}}{\partial \textbf{l}}-\textbf{m}\times\frac{\partial \mathcal{H}}{\partial \textbf{l}},
\end{align}
\end{subequations}
where at equilibrium, $\textbf{m}=\textbf{0}$ and $\textbf{l}=2S\hat{\textbf{z}}$. Thus, the linearized equations of motion yield
\begin{subequations}
\begin{align}
\hbar\frac{d\textbf{m}}{dt} &= 2SD_yl_y \hat{\textbf{x}}-2SD_xl_x \hat{\textbf{y}}, \\ 
\hbar\frac{d\textbf{l}}{dt} &= 2S(J+D_y)m_y \hat{\textbf{x}}-2S(J+D_x)m_x \hat{\textbf{y}},
\end{align}
\end{subequations}
from which we deduce that
\begin{subequations}
\begin{align}
    \frac{d^2m_x}{dt^2}+\frac{4S^2}{\hbar^2}D_y(J+D_x)m_x&=0, \\
    \frac{d^2l_y}{dt^2}+\frac{4S^2}{\hbar^2}D_y(J+D_x)l_y&=0, \\
     \frac{d^2m_y}{dt^2}+\frac{4S^2}{\hbar^2}D_x(J+D_y)m_y&=0, \\
      \frac{d^2l_x}{dt^2}+\frac{4S^2}{\hbar^2}D_x(J+D_y)l_x&=0. 
\end{align}
\end{subequations}
We therefore obtain the two eigenfrequencies as
\begin{subequations}
\begin{align}
\hbar\Omega_1&=2S\sqrt{D_x(J+D_y)}, \\
\hbar\Omega_2&=2S\sqrt{D_y(J+D_x)}.
\end{align}
\end{subequations}

\subsection*{C.2 Perturbative solutions of the Landau-Lifshitz-Gilbert equations}

The Landau-Lifshitz-Gilbert equations are given by
\begin{equation}
\frac{d \textbf{S}_\sigma}{dt} =  \frac{\gamma_{el}}{1+\kappa_{el}^2}\left[ \textbf{S}_\sigma \times \textbf{B}_\sigma^{\text{eff}} -\frac{\kappa_{el}}{|\textbf{S}_\sigma|} \textbf{S}_\sigma \times \left( \textbf{S}_\sigma \times \textbf{B}_\sigma^{\text{eff}}  \right) \right], \\
\label{eq:LLGeqs}
\end{equation}
where the effective magnetic field can be obtained by
\begin{equation}
\begin{split}
\textbf{B}_\sigma^{\text{eff}} &= -\frac{1}{\hbar\gamma_{el}} \frac{\partial \mathcal{H}}{\partial \textbf{S}_\sigma} \\
&= \textbf{B} -\frac{1}{\hbar\gamma_{el}}\left(  J \textbf{S}_{\sigma'}+2D_xS_{\sigma,x} \hat{\textbf{x}} + 2D_y S_{\sigma,y} \hat{\textbf{y}} \right).
\end{split}
\end{equation}
Here, $\mathbf{B}$ is the effective magnetic field created by the driving force of the exciting particles. $\sigma,\sigma' \in \{1,2\}$, with $\sigma \neq \sigma'$, and we set $|\textbf{S}_\sigma|=S$. Using
\begin{equation}
\begin{split}
\textbf{S}_\sigma \times\left( \textbf{S}_\sigma \times \textbf{B}_\sigma^{\text{eff}} \right) &= (\textbf{S}_\sigma \cdot \textbf{B}_\sigma^{\text{eff}}) \textbf{S}_\sigma - \left|  \textbf{S}_\sigma \right|^2 \textbf{B}_\sigma^{\text{eff}} \\ &= (\textbf{S}_\sigma \cdot \textbf{B}_\sigma^{\text{eff}}) \textbf{S}_\sigma - S^2 \textbf{B}_\sigma^{\text{eff}},
\end{split}
\end{equation}
we obtain
\begin{align}
\frac{d \textbf{S}_\sigma}{dt} = & \frac{\gamma_{el}}{1+\kappa_{el}^2} \nonumber\\
& \times \left[ \textbf{S}_\sigma \times \textbf{B}_\sigma^{\text{eff}} -\frac{\kappa_{el}}{S} (\textbf{S}_\sigma \cdot \textbf{B}_\sigma^{\text{eff}}) \textbf{S}_\sigma +\kappa_{el} S\textbf{B}_\sigma^{\text{eff}} \right].
\end{align}
We assume the magnetic field to be aligned along the $x$-axis, $\textbf{B} \equiv B\hat{\textbf{x}}$. The Hamiltonian is invariant under the exchange of $\textbf{S}_1 \leftrightarrow \textbf{S}_2$. Furthermore, since we have an antiferromagnetic system ($J>0$), we conclude that $S_{1,y}=-S_{2,y}$ and $S_{1,z}=-S_{2,z}$. However, the interaction between $\textbf{B}$ and $\textbf{S}$ enforces $ S_{1,x} = S_{2,x}$. With these considerations, we can write the effective magnetic field as

\begin{align}
{\textbf{B}_1^{\text{eff}}} = &  \left( B - \frac{J+2D_x}{\hbar\gamma_{el}}S_{1,x} \right) \hat{\text{x}} \nonumber\\
& + \frac{J-2D_y}{\hbar\gamma_{el}} S_{1,y} \hat{\textbf{y}} + \frac{J}{\hbar\gamma_{el}} S_{1,z}  \hat{\textbf{z}}.
\end{align}
The cross product with the spins therefore yields
\begin{equation}
\begin{split}
\textbf{S}_1 & (t)  \times {\textbf{B}_1^{\text{eff}}} (t) = \frac{2D_y}{\hbar\gamma_{el}} S_{1,y}(t)S_{1,z}(t) \hat{\textbf{x}} \\
&+ \left( S_{1,z}(t)B(t) - \frac{2(J+D_x)}{\hbar\gamma_{el}}S_{1,x}(t)S_{1,z}(t) \right) \hat{\textbf{y}} \\
&+\left( \frac{2(J+D_x-D_y)}{\hbar\gamma_{el}} S_{1,x}(t)S_{1,y}(t)-S_{1,y}(t)B(t) \right) \hat{\textbf{z}}.
\end{split}
\end{equation}

To solve the Landau-Lifshitz-Gilbert equations up to first nonzero order in the external electric field, we Fourier transform them to the frequency domain. We perform an expansion of each field $A(\omega)$ of the form $A(\omega)=A^{(0)}(\omega) + A^{(1)}(\omega) + A^{(2)}(\omega) + \cdots $, where $A^{(k)}(\omega) \propto E_0^k$ and $E_0$  is the amplitude of the electric field. 
We assume the effective magnetic field induced by exciting particles to be in second order of the electric field component, $B(\omega) = B^{(2)}(\omega) + O(E_0^3)$. The spins are initially aligned along the $z$-axis. Therefore, $S_{1,x}(\omega) = S_{1,x}^{(2)}(\omega) + O(E_0^3) $, $S_{1,y}(\omega) =S_{1,y}^{(2)}(\omega) +O(E_0^3)$, and $S_{1,z}(\omega)=  \sqrt{2\pi}S\delta(\omega) + S_{1,z}^{(4)}(\omega) + O(E_0^5) $. Note that $X(\omega) \circledast S_{1,z}^{(1)}(\omega) = S X(\omega)$, where $\circledast$ stands for convolution. Thus,
\begin{subequations}
\begin{align}
\begin{split}
i\omega {S_{1,x}^{(2)}} (\omega) &= \frac{\gamma_{el}}{1+\kappa_{el}^2} \left[ \frac{2SD_y}{\hbar\gamma_{el}} {S_{1,y}^{(2)}}(\omega) -\frac{\kappa_{el}SJ}{\hbar \gamma_{el} } {S_{1,x}^{(2)}}(\omega) \right. \\ & \left. + S\kappa_{el}\left( B^{(2)}(\omega) 
- \frac{J+2D_x}{\hbar\gamma_{el}} {S_{1,x}^{(2)}} (\omega) \right) \right], 
\end{split} \\
\begin{split}
i\omega {S_{1,y}^{(2)}}(\omega) &= \frac{\gamma_{el}}{1+\kappa_{el}^2} \left[ SB^{(2)}(\omega) - \frac{2S(J+D_x)}{\hbar\gamma_{el}} {S_{1,x}^{(2)}}(\omega) \right. \\ 
& \left. - \frac{\kappa_{el}SJ}{\hbar\gamma_{el}}{S_{1,y}^{(2)}}(\omega) + \kappa_{el}S \frac{J-2D_y}{\hbar\gamma_{el}} {S_{1,y}^{(2)}}(\omega)  \right], 
\end{split} \\
\begin{split}
i \omega {S_{1,z}^{(4)}}(\omega) &= \frac{\gamma_{el}}{1+\kappa_{el}^2} \left[ \frac{2(J+D_x-D_y)}{\hbar\gamma_{el}} {S_{1,x}^{(2)}}(\omega) \circledast {S_{1,y}^{(2)}}(\omega) \right. \\
&- {S_{1,y}^{(2)}}(\omega) \circledast B^{(2)}(\omega)  - \kappa_{el}\left( {S_{1,x}^{(2)}}(\omega) \circledast B_\text{ph}^{(2)}(\omega) \right. \\
&  +\frac{J-2D_y}{\hbar\gamma_{el}} {S_{1,y}^{(2)}}(\omega) 
 \circledast {S_{1,y}^{(2)}}(\omega) \\
 &\left. - \frac{J+2D_x}{\hbar\gamma_{el}} {S_{1,x}^{(2)}}(\omega) \circledast {S_{1,x}^{(2)}}(\omega)  \right)   \\
 &-\left. \frac{2 S\kappa_{el} J}{\hbar\gamma_{el}} {S_{1,z}^{(4)}}(\omega) \right] .
\end{split}
\end{align}
\end{subequations}

The above equations have the following solutions up to first nonzero order in the external electric field:
\begin{subequations}
\begin{align}
 S_{1,x}(\omega) =& \frac{\gamma_{el} S}{1+\kappa_{el}^2} \frac{i\kappa_{el}\omega+2SD_y/\hbar}{\Omega_m^2-\omega^2+i\kappa_m\omega} B^{(2)}(\omega), \\
 S_{1,y}(\omega) =& \frac{\gamma_{el}S}{1+\kappa_{el}^2} \frac{i\omega}{\Omega_m^2-\omega^2+i\kappa_m\omega} B^{(2)}(\omega), \\
 S_{1,z}(\omega) =&  \sqrt{2\pi} S\delta(\omega) + \frac{\gamma_{el}}{i\omega(1+\kappa_{el}^2)+\kappa_{el}2S J/\hbar} \nonumber \\
 & \cdot \left[\kappa_{el} S_{1,x}(\omega) \circledast \left( \frac{J+2D_x}{\hbar\gamma_{el}}S_{1,x}(\omega)  \right.\right. \nonumber \\
 & \left. - B^{(2)}(\omega) \right)   + \left( \frac{2(J+D_x-D_y)}{\hbar\gamma_{el}} S_{1,x}(\omega) \right. \nonumber \\
  &\left. -\kappa_{el}\frac{J-2D_y}{\hbar\gamma_{el}} S_{1,y}(\omega) -B^{(2)}(\omega)\right) \circledast  \nonumber \\
  &  S_{1,y}(\omega)  \Bigr],
\end{align}
\end{subequations}
where $\Omega_m = \frac{2S}{\hbar}\sqrt{\frac{(J+D_x)D_y}{1+\kappa_{el}^2}}$ and $\kappa_m=\frac{2S\kappa_{el}}{\hbar\left(1+\kappa_{el}^2\right)}(J+D_x+D_y)$. Hence, the magnetic field excites the high-frequency magnon.

We obtain the rectified components by evaluating the factor that accompanies $B^{(2)}$ at $\omega=0$,
\begin{subequations}
\begin{align}
\langle{S_{1,x}}\rangle(t) &= \frac{\gamma_{el}\hbar}{2(J+D_x)} B^{(2)}(t), \\
 \langle{S_{1,y}}\rangle(t) &= 0, \\
 \langle{S_{1,z}}\rangle(t) &= S - \frac{\gamma_{el}^2\hbar^2}{8S(J+D_x)^2}  {B^{(2)}}^2(t).
\end{align}
\end{subequations}
The energy is therefore given by

\begin{equation}
\mathcal{H}(t) = -J  - \frac{\gamma_{{el}}^2\hbar^2}{2(J+D_x)} {B^{(2)}}^2(t) ,
\end{equation}
whereas the magnetization can be expressed as 
\begin{equation}
M_x(\omega) = \frac{2\hbar\gamma_{el}}{V_c}S_{1,x}(\omega) .
\label{eq:Mx_frequency}
\end{equation}

Let
\begin{align}
 \mathcal{T}_m(t) =& \sqrt{2\pi}e^{-\kappa_S t/2}\frac{\gamma_{el}S\kappa_{el}}{1+\kappa_{el}^2} \nonumber \\
 & \cdot \left[\cos\left(\widetilde{\Omega}_m t\right)+\frac{4S D_y-\kappa_{el}\kappa_m\hbar}{2\kappa_{el}\widetilde{\Omega}_m\hbar}\sin(\widetilde{\Omega}_m t) \right]\theta(t), 
\end{align}
where $\widetilde{\Omega}_m = \sqrt{\Omega_m^2-\frac{\kappa_m^2}{4}}$. Then
\begin{equation}
    S_{1,x}(t) = \mathcal{T}_m(t) \circledast B(t),
\label{eq:Sx_time}
\end{equation}
from which we can calculate the time-dependent magnetization as
\begin{equation}
    M_x(t) = \frac{2\hbar\gamma_{el}}{V_c}S_{1,x}(t) .
\label{eq:Mx_time_domain}
\end{equation}

\begin{subequations}
\begin{align}
Q_y(t) &= \frac{I^{(2)}_y(t)}{\widetilde{\Omega}_y}, \\
\dot{Q}_y(t) &= I^{(1)}_y(t) -\frac{\kappa_y}{2\widetilde{\Omega}_y}I^{(2)}_y(t), \\
Q_z(t) &= \frac{I^{(4)}_z(t)}{\widetilde{\Omega}_z}, \\
\dot{Q}_z(t) &= I^{(3)}_z(t) -\frac{\kappa_z}{2\widetilde{\Omega}_z}I^{(4)}_z(t),
\end{align}
\end{subequations}
allows us to find expressions for the effective magnetic field $B(t)$ of each mechanism. Here, $\widetilde{\Omega}_\alpha = \sqrt{\Omega_\alpha^2-\frac{\kappa_\alpha^2}{4}}$ for phonons $\alpha$, $\alpha \in \{ y,z\}$. $I_{\alpha}^{(\sigma)}$, $\sigma\in\{1,2,3,4\}$, refers to any of the functions in Eqs.~\eqref{eq:integralsI}, after replacing the parameters $\Omega$ and $\kappa$ with those of the respective phonon $\alpha$, and $E_0$ and $\tau$ with those of the circularly polarized electric field used in the excitation.

Convolutions can be done efficiently in Python with NumPy's \texttt{convolve} function.


\begin{figure}[t]
\centering
\includegraphics[scale=0.61]{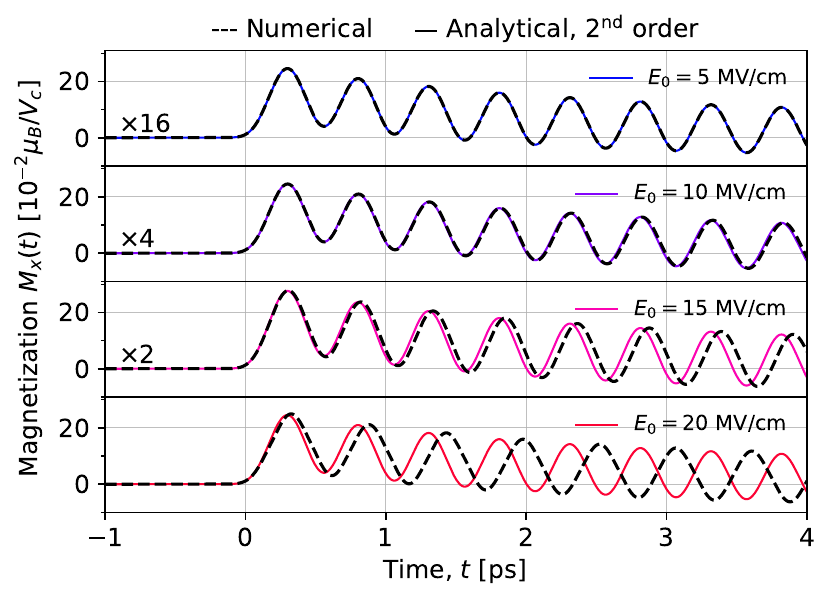}
\caption{\textbf{Comparison of analytical and numerical results.} 
  Magnon dynamics as evaluated by the semi-analytical expression in Eq.~\eqref{eq:Sx_time} compared to the numerical evaluations of the equation of motion Eq. ~\eqref{eq:LLGeqs}. We show the example of two-phonon excitation, using different amplitudes of the peak electric field. By increasing the laser intensity, it becomes evident that higher-order corrections are necessary. In this work, we only discuss contributions up to second order, and we use parameters for which the numerical results agree well with the analytical ones. 
  }
\label{fig:analyticalcomparisonmagnons}
\end{figure}

\begin{figure}[h]
    \centering
    
    
      
      
      
    
    \includegraphics[width=0.47\textwidth]{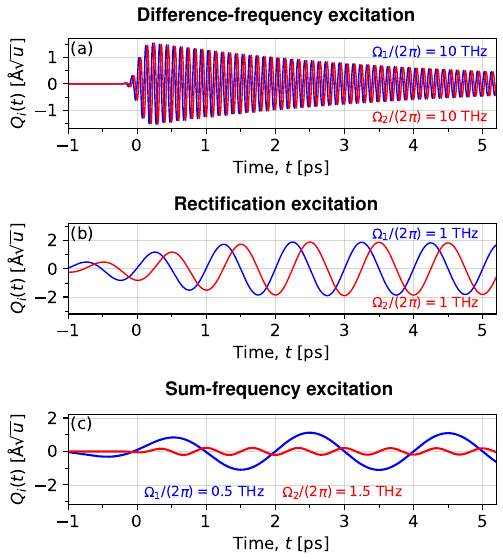}
    \caption{
    Amplitude $Q_i$, $i \in \{1,2\}$, of the infrared-active phonon modes resonantly excited by the laser pulse, participating in the driving force of the two-phonon and one-photon-one-phonon magneto-phononic mechanisms. (a) Phonon amplitude for the IRS and IRRS mechanisms in Fig.~\ref{fig:main2}(a). (b) Phonon amplitude for the corresponding rectification mechanisms in Fig.~\ref{fig:main2}(d). (c) Phonon amplitude for the SF-IRS and SF-IRRS mechanisms in Fig.~\ref{fig:main2}(g).
    }
    \label{fig:Qd2}
\end{figure}

\clearpage{}



%

\end{document}